\documentclass[nofootinbib,twocolumn,aps,prd,amsmath,superscriptaddress]{revtex4-1}
\usepackage{setspace}
\usepackage{color}
\usepackage{fancyhdr}
\usepackage{graphicx}
\usepackage[ansinew]{inputenc}
\usepackage{amssymb}
\usepackage{amsmath}
\usepackage{cancel}
\usepackage{float}
\usepackage{graphicx}
\usepackage{float}
\usepackage{subfigure}
\usepackage{cancel}
\usepackage{tabularx}
\usepackage{cleveref}

\newcommand{\rom}[1]{\MakeUppercase{\romannumeral #1}}
\begin{document}
\title{Twin stars and the stiffness of the nuclear equation of state:
	ruling out strong phase transitions below $1.7n_0$ with the new NICER
	radius measurements }
\author{Jan-Erik Christian}
\email{christian@astro.uni-frankfurt.de}
\affiliation{Institut f\"ur Theoretische Physik, Goethe Universit\"at Frankfurt, 
	Max von Laue Strasse 1, D-60438 Frankfurt, Germany}


\author{J\"urgen Schaffner-Bielich}
\email{schaffner@astro.uni-frankfurt.de}
\affiliation{Institut f\"ur Theoretische Physik, Goethe Universit\"at Frankfurt, 
Max von Laue Strasse 1, D-60438 Frankfurt, Germany}
\date{\today}

  \begin{abstract}
	We explore the connection between the stiffness of an hadronic equation of state (EoS) with a sharp phase transition to quark matter to its tidal deformability. For this we employ a hadronic relativistic mean field model with a parameterized effective nucleon mass to vary the stiffness in conjunction with a constant speed of sound EoS for quark matter. We compute multiple scenarios with phase transitions according to the four possible cases of a hybrid star EoS with a stable second branch. We demonstrate at the example of GW170817 how the effective nucleon mass can be constrained by using gravitational wave data. We find, that certain values of the effective nucleon mass are incompatible with GW170817 and a phase transition simultaneously. By using the recent NICER measurements of J0030+0451 at the $1\sigma$ level we constrain our results further and find that strong phase transitions with a visible jump in the mass-radius relation are ruled out at densities below 1.7 times saturation density.
  \end{abstract}
\maketitle
\section{Introduction}\label{incanus}
 A well established hypothesis for the equation of state of compact stars is the possibility of hybrid stars \cite{Ivanenko:1965dg,Itoh:1970uw,Alford:2004pf,Coelho:2010fv,Chen:2011my,Masuda:2012kf,Yasutake:2014oxa,Zacchi:2015oma}, which feature a hadronic mantle and a quark matter core. In contrast to pure hadronic EoSs, which generate a single stable branch in a mass-radius relation, these hybrid EoSs can generate a second stable branch. This can lead to so called twin stars, where two stars have the same mass, but different radii \cite{Kampfer:1981yr,Glendenning:1998ag,Schertler:2000xq,SchaffnerBielich:2002ki,Zdunik:2012dj,Alford:2015dpa,Blaschke:2015uva,Zacchi:2016tjw,Alford:2017qgh,Christian:2017jni,Blaschke:2019tbh}.\\
The possibility of pure quark stars is not ruled out either \cite{Ivanenko:1965dg,Itoh:1970uw,Bodmer:1971we,Haensel:1986qb,Alcock:1986hz,Fraga:2001xc,Zacchi:2015lwa}. A widely used approach to describe the hadronic matter in neutron star is the relativistic mean field model \cite{PhysRev.98.783,Duerr56,Walecka74,Boguta:1977xi,Serot:1984ey,Mueller:1996pm,Typel:2009sy,Hornick:2018kfi}.\\ 
The chirp mass $\mathcal{M}$ and the weighted tidal deformability $\tilde{\Lambda}$ can be measured from the inspiral of two neutron stars. This makes gravitational wave data useful in constraining the EoS for neutron stars \cite{TheLIGOScientific:2017qsa,Annala:2017llu,Bauswein:2017vtn,Paschalidis:2017qmb}. 
Specifically, hybrid star EoS, due to their high compactness, fit well with the low values of tidal deformability measured for GW170817 \cite{Paschalidis:2017qmb,Alvarez-Castillo:2018pve,Christian:2018jyd,Montana:2018bkb,Sieniawska:2018zzj}. Another important constraint is the maximal observed mass of a neutron star. Currently the highest measured mass for a neutron star is about $2M_\odot$ \cite{Demorest:2010bx,Antoniadis:2013pzd,Fonseca:2016tux} or slightly higher at  $2.14_{-0.09}^{+0.10}M_\odot$ \cite{Cromartie:2019kug}.\\
In the light of the recent first measurement of a neutron star merger (GW170817 \cite{TheLIGOScientific:2017qsa}) we explore the influence of the stiffness and transition parameters of a hadronic EoS featuring a first order phase transition to quark matter. To this end we employ the parameterizable relativistic mean field equation of state by Hornick et al. \cite{Hornick:2018kfi}, which enables us to vary the effective nucleon mass. The effective nucleon mass is linked to the stiffness of the EoS \cite{1983PhLB..120..289B}, see also Yasin et al. \cite{Yasin:2018ckc}. The phase transition and quark matter EoS is modeled after the constant speed of sound parametrization presented by Alford et al. \cite{Alford:2014dva}. The parameters for the phase transition are chosen according to the four categories of twin stars outlined in \cite{Christian:2017jni}. We find, that the presence of a phase transition can have significant influence on the compatibility of the underlying hadronic EoS with the GW170817 data, making EoSs previously considered to be too stiff viable. However, a soft EoS might not be capable of generating a second branch in the mass-radius relation. This way certain assumptions of a phase transition for a known stiffness of the nuclear EoS can be excluded. The recent measurements by NICER \cite{Riley:2019yda,Miller:2019cac,Raaijmakers:2019qny} of the pulsar J0030+0451 can be used to constrain the EoS further. Riley at al. state a mass of $1.34^{+0.15}_{-0.16}M_\odot$ with a radius of $12.71^{+1.14}_{-1.19}\mathrm{km}$ \cite{Riley:2019yda}, while Miller et al. state $1.44^{+0.15}_{-0.14}M_\odot$ with a radius of $13.02^{+1.24}_{-1.06}\mathrm{km}$ \cite{Miller:2019cac}. 
This constraint rules out a strong phase transtion at densities of  $n \lesssim 1.7\,n_0$.  We show that the NICER data \cite{Riley:2019yda,Miller:2019cac,Raaijmakers:2019qny} provides an indication, that an extremely soft nuclear equation of state and a strong phase transition are mutually exclusive.
\section{Theoretical Framework}
\subsection{Equation of State} 
\subsubsection{Hadronic Equation of State} 
The relativistic parametrization introduced by Todd-Rudel et al. \cite{ToddRutel:2005fa} (see also: \cite{Chen:2014sca,Hornick:2018kfi}) is a generalized relativistic mean field approach with the main advantage, that the slope parameter $L$, the symmetry energy $J$ and the effective nucleon mass $m^*/m$ can be easily adjusted. Taking into account $\sigma$, $\omega$ and $\rho$ mesons, the interaction Lagrangian can be written as: 
\begin{align}\label{Eq:Nadine_Lagrangian}
\begin{aligned} \mathcal { L } _ { \mathrm { int } } & = \sum _ { N } \overline { \psi } _ { i } \left[ g _ { \sigma } \sigma - g _ { \omega } \gamma ^ { \mu } \omega _ { \mu } - \frac { g _ { \rho } } { 2 } \gamma ^ { \mu } \vec { \tau } \vec { \rho } _ { \mu } \right] \psi _ { i } \\ & - \frac { 1 } { 3 } b m \left( g _ { \sigma } \sigma \right) ^ { 3 } - \frac { 1 } { 4 } c \left( g _ { \sigma } \sigma \right) ^ { 4 } \\ & + \Lambda _ { \omega } \left( g _ { \rho } ^ { 2 } \vec { \rho } _ { \mu } \vec { \rho } ^ { \mu } \right) \left( g _ { \omega } ^ { 2 } \omega _ { \mu } \omega ^ { \mu } \right) + \frac { \zeta } { 4 ! } \left( g _ { \omega } ^ { 2 } \omega _ { \mu } \omega ^ { \mu } \right) ^ { 2 } \end{aligned}
\end{align}
The last two terms describe a density dependence via the $\sigma-\omega$ coupling term $\Lambda_{\omega}$ and the quadratic self coupling $\zeta$ of the $\omega$ mesons \cite{Mueller:1996pm,Horowitz:2001ya,ToddRutel:2005fa}.
The $g_\sigma$ and $g_\omega$ couplings can be used to determine the density of the ground state $n_0$, as well as the binding energy per particle $E/A(n_0)$.\\
If one wants to determine the values of $E/A(n_0)$, $b$, $c$ and $\Lambda_{ \omega }$ one needs to fix certain parameters. Hornick et al. \cite{Hornick:2018kfi} followed the approach by  Chen et al. \cite{Chen:2014sca} to do so. 
Apart from $n_0$, $E/A(n_0)$, incompressibility $K(n_0)$, the parameters $J$, $L$ and $m^*/m$ have to be fixed. The value of $\zeta$ is set to zero \cite{Chen:2014sca} in the following to achieve the stiffest possible EoS. $K$ is fixed to $K=20\mathrm{MeV}$ \cite{Hornick:2018kfi}. Hornick et al. additionally constrain the fixed parameters using the constraints from the analysis of $\chi\mathrm{EFT}$ for densities up to $1.3\,n_0$ \cite{Drischler:2016djf}. By comparing the different EoSs with the allowed band from $\chi\mathrm{EFT}$ they find, that only values of $40\le L \le 60$ are possible, when $30\le J \le 32$ also holds true.\\
We fixed the values $L=60\,\mathrm{MeV}$ and $J=32\,\mathrm{MeV}$ while varying the effective mass $m^{*}/m$. These values of $L$ and $J$ allow for the greatest allowed range in effective mass values, see \cite{Hornick:2018kfi}. We note that the mass-radius relation does not depend significantly on the choices of $L$ and $J$ \cite{Hornick:2018kfi}. The softness of an EoS corresponds to the value of $m^{*}/m$, as only $m^*/m$ controls the high-density behavior \cite{1983PhLB..120..289B}. Lower values of $m^{*}/m$ generate a softer EoS, while high values generate a stiffer EoS. 

\subsubsection{Phase Transition}
We assume, that at high baryonic densities a first order phase from hadronic to quark matter takes place. This behavior is modeled with a Maxwell construction. The hadronic matter is described by the parameterized EoS (see \cite{Hornick:2018kfi}), while the constant speed of sound approach \cite{Zdunik:2012dj,Alford:2014dva,Alford:2015gna} in the form used by Alford et al. \cite{Alford:2014dva} is employed for the quark matter. 
This means, the entire EoS is given as:
\begin{equation}
	\epsilon(p) =
	\begin{cases} 
		\epsilon_{HM}(p)	&  p < p_{trans}\\
		\epsilon_{HM}(p_{trans})+\Delta\epsilon + c_{QM}^{-2}(p-p_{trans})	& p > p_{trans}\\
	\end{cases}
\end{equation} 
where $p_{trans}$ is the pressure at which the transition takes place and $\epsilon$ the corresponding energy density.
The discontinuity in energy density at the transition is $\Delta\epsilon$. 
For the speed of sound in the stars core, a value of $c_{QM}=1$ is assumed, using natural units.

\subsection{Classification of Twin Stars}\label{Tevildo}
A first order phase transition gives rise to the phenomenon of "twin stars", which are neutron stars with identical mass, but different radii \cite{Glendenning:1998ag,Schertler:2000xq,SchaffnerBielich:2002ki,Zdunik:2012dj,Alford:2015dpa,Blaschke:2015uva,Zacchi:2016tjw,Christian:2017jni}. In order to investigate twin star equations of state it can be useful to classify the twin star solutions into four distinct categories, as described in \cite{Christian:2017jni}. In this subsection a short summary of the four categories is provided. We refer to the maximum of the hadronic branch as the first maximum and the maximum of the hadronic branch as the second maximum in a twin star mass-radius relation.
In \cite{Christian:2017jni} we showed that the mass value of the first and second maximum can be related to values of $p_{trans}$ and $\Delta\epsilon$ respectively. The shape of the second branch is governed by the value of $p_{trans}$, while its position is strongly influenced by the value of $\Delta\epsilon$. High values of $p_{trans}$ lead to high masses in the first maximum and flat second branches. Low values of $\Delta\epsilon$ lead to a second branch near the discontinuity (i.e. a high mass at the second maximum). With this in mind the twin star categories can be defined as follows:
\begin{itemize}
	\item [\textbf{\rom{1}:}] Both maxima exceed $2M_\odot$, which implies high values of $p_{trans}$ and a nearly flat second branch.  
	\item [\textbf{\rom{2}:}] Only the first maximum reaches $2M_\odot$, which again requires a high value of $p_{trans}$.
	\item [\textbf{\rom{3}:}]The first maximum is in the range of $2M_\odot \geq M_{max_1} \geq 1M_\odot$, while the second maximum exceeds $2M_\odot$. Accordingly, the transitional pressure is lower than in the previous categories and the second branch becomes steeper. 
	\item [\textbf{\rom{4}:}] Like category \rom{3}  the second maximum exceeds $2M_\odot$, however the first maximum is below even $1M_\odot$. The second branch is at its steepest slope here. 
\end{itemize}
\subsection{Tidal deformability}\label{gandalf}
The observation of gravitational waves from compact star mergers, as demonstrated for GW170817 detected by the LIGO and Virgo
observatories \cite{TheLIGOScientific:2017qsa}, can be used to constrain
the EoSs of compact stars, because they contain information on the tidal deformability and chirp mass of the participating neutron stars during the inspiral phase.\\
The chirp mass can be measured to a very high precision and is closely related to the total mass $M_{total}$ via:
\begin{equation}
\mathcal{M}=\left(\frac{q}{(1+q)^2}\right)^{\frac{3}{5}}M_{total}
\end{equation}
where $q$ is the mass-ratio of the participating stars. For GW170817  the chirp mass was measured as $\mathcal{M}=1.186^{+0.001}_{-0.001}M_\odot$ \cite{Abbott:2018wiz}.\\
The tidal deformability $\lambda$ measures the quadrupole
deformation $Q_{ij}$ of an object in response to the external tidal field
$\mathcal{E}_{ij}$ \cite{Hinderer:2007mb,Hinderer:2009ca} in the following form:
\begin{equation}
	Q_{ij}=-\lambda \mathcal{E}_{ij} 
\end{equation}
where $\lambda$ is related to the more commonly used parameter $\Lambda$ in the following way:
 \begin{equation}
 \Lambda=\frac{2k_2}{3C^5} 
 \end{equation}
with $k_2=\frac{3}{2}\lambda R^{-5}$ and $C=M/R$ being the compactness of the star.\\
The most interesting aspect of $\Lambda$ for our purposes is, that it is dependent on the EoS of the compact star, that is being deformed, and that it can be easily calculated \cite{Hinderer:2007mb,Hinderer:2009ca,Postnikov:2010yn}. This can be used to compare the calculated values with the gravitational wave measurement. However, the inspiral of two compact stars with masses $M_1\ge M_2$ can only reveal a combined value of the tidal deformabilities $\tilde{\Lambda}$. For this reason $\Lambda_1-\Lambda_2$ plots are common, where every value of $\Lambda_1$ is assigned a fitting value of $\Lambda_2$ based on the precisely measured chirp mass. Depending on the EoS this can lead up to three thin lines in the plot. These lines are a neutron-neutron (NN), neutron-hybrid (NH) and hybrid-hybrid star line (HH) (see for more detail: refs. \cite{Christian:2018jyd,Montana:2018bkb}). Each dot in these plots indicates a possible pair of merging neutron stars. The gravitational wave data can then be used to constrain the area in the $\Lambda_1-\Lambda_2$ plot from which the measured signal would have originated. 
Due to the high mass values of all hybrid stars in category \rom{1} the LIGO measurement excludes the participation of a category \rom{1} hybrid star in the GW170817 event (see \cite{Christian:2018jyd}). Category \rom{1} EoSs might still be viable. However, using GW170817 data they are identical to the purely hadronic case and will thus not be discussed separately. 
\section{Tidal deformability from equations of state with varying stiffness}
\begin{figure}
	\centering				
	\includegraphics[width=8.5cm]{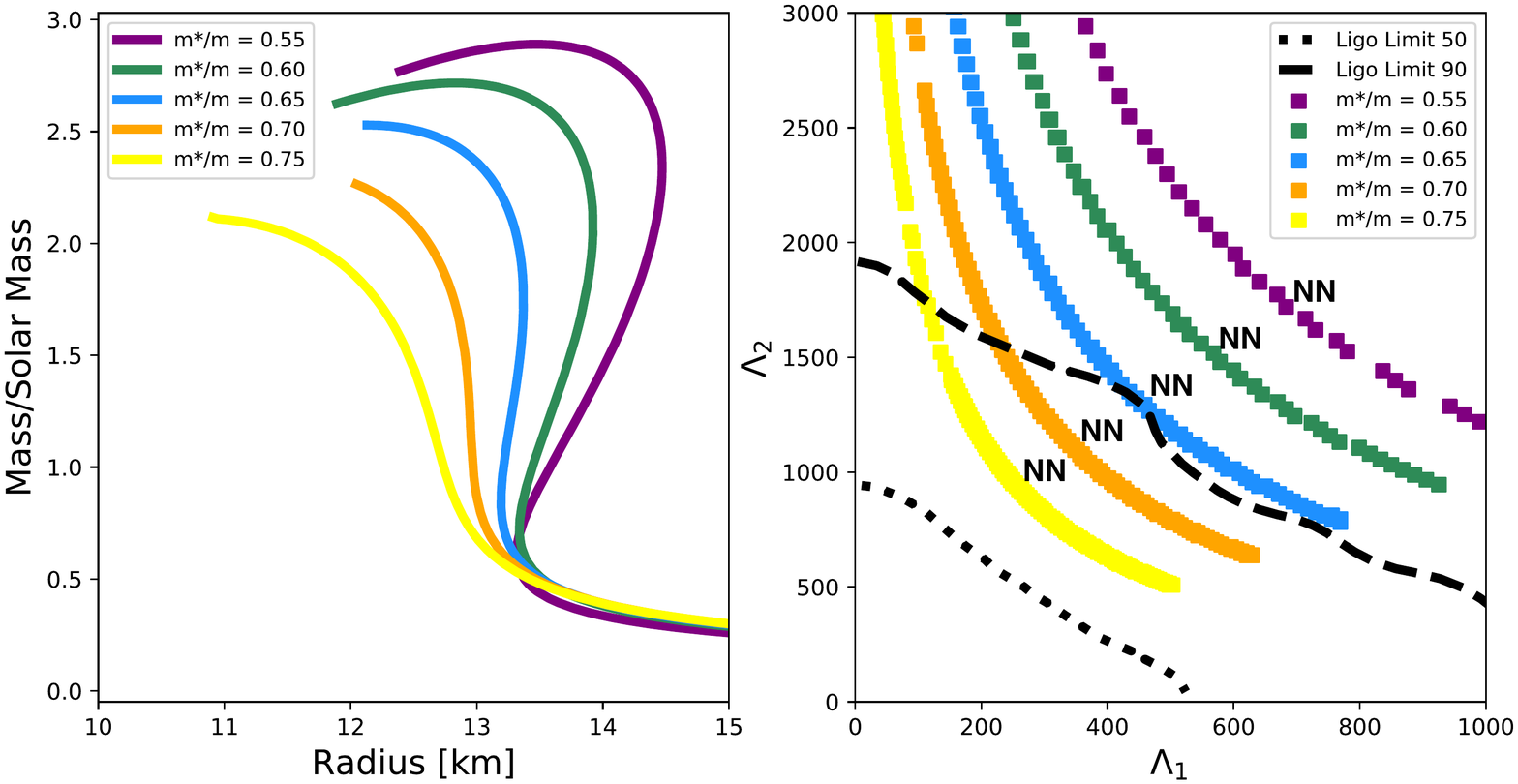}
	\caption{\footnotesize On the left side the mass-radius relation for an EoS with $J = 32\,MeV$ and $L = 60\,MeV$, with varied values of $m^*/m$ is displayed. On the right side the corresponding possible neutron star combinations are shown. }
	\label{LvLCatpure}
\end{figure}

\begin{figure}
	\centering				
	\includegraphics[width=8.5cm]{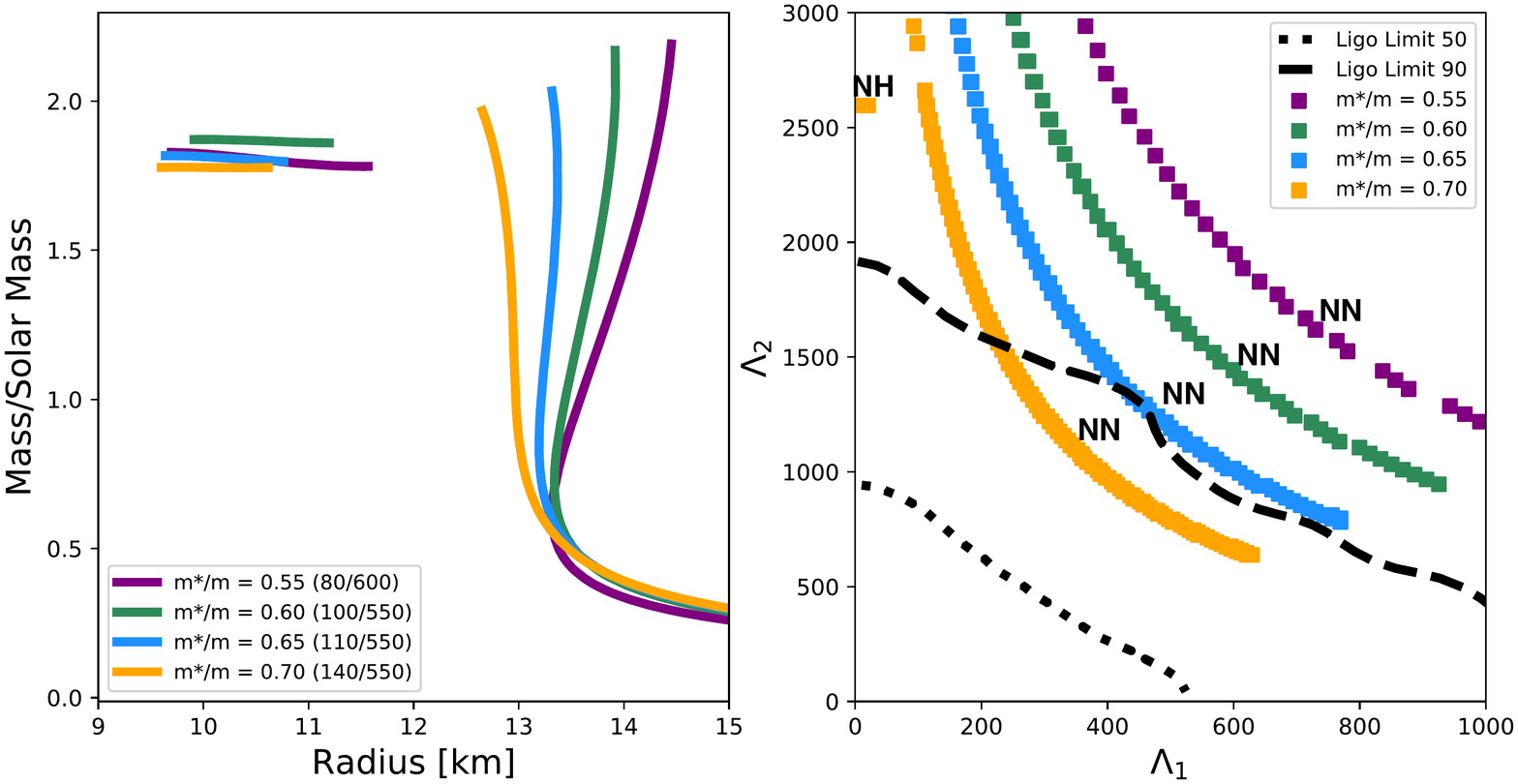}
	\caption{\footnotesize A phase transition of the category II type is depicted, the parameters are written behind the corresponding values of $m^*/m$ in the order ($p_{trans}$/$\Delta\epsilon$) in units of $\mathrm{MeV/fm^3}$. The NN combinations are identical to the pure case, due to the late phase transitions. NH combinations close to the axis can be found for all cases. However, the NH combinations are not closer to the LIGO limit, than the NN combinations.}
	\label{LvLCatII}
\end{figure}

In the following we present the $\Lambda_{1}-\Lambda_{2}$ plots for a selection of EoSs from the categories \rom{2} - \rom{4}, as well as the pure hadronic case, described in \cite{Christian:2017jni}, with varying $m^*/m$ using the chirp mass and credibility limits from GW170817 as constraints. 
The effective mass starts at $m^*/m = 0.55$ and is increased in steps of $m^*/m=0.05$ to $m^*/m=0.75$. A slope parameter of $L=60\,\mathrm{MeV}$ and a symmetry energy of $J=32\,\mathrm{MeV}$ are fixed. We start with the pure hadronic case in figure \ref{LvLCatpure}. On the left plot are the mass-radius relations and on the right one are the corresponding $\Lambda_{1}-\Lambda_{2}$ plots. The 90\% and 50\% credibility levels by LIGO are added into the graphic as a dashed and a dotted black line respectively \cite{Abbott:2018wiz}. Like Hornick et al. \cite{Hornick:2018kfi} we find, that effective masses of $m^*/m \ge 0.65$ are compatible with GW170817 data.\\
Ideally one would keep the parameters of $p_{trans}$ and $\Delta\epsilon$ identical for all variations of $m^*/m$ within a category, in order to investigate the effect of a varied stiffness in isolation. However, in order to find category \rom{2} solutions high values of $p_{trans}$ and $\Delta\epsilon$ are necessary and due to the high transitional pressure it is not possible to find a single value of $p_{trans}$ that can generate a phase transition for all investigated values of $m^*/m$. For this reason the $p_{trans}$ and $\Delta\epsilon$  parameters are chosen to be as close together as possible while still generating a category \rom{2} solution. For a hadronic EoS as soft as the $m^*/m=0.75$ case it is not possible to find a category \rom{2} solution at all.
The mass radius relations (left) and the $\Lambda_{1}-\Lambda_{2}$ plots (right) from category \rom{2} are shown in figure \ref{LvLCatII}.\\ 
Only $m^*/m=0.65$ and $m^*/m=0.70$ generate NN pairs within the credibility limit, as is the case in the purely hadronic scenario. For all category \rom{2} EoSs the NH pairs are close to the y-axis. This is caused by the high mass values of the hybrid stars in this category. Stiffer hadronic EoSs seem to generate their corresponding NH pairs at higher values of $\Lambda_{2}$. However, even for the softest EoS with $m^*/m = 0.70$ the NH pairs are still above the 90\% credibility level. This means, that the compatibility of a category \rom{2} EoS with GW170817 depends entirely on the hadronic EoS, since only NN combinations are within the LIGO credibility level.

\begin{figure}
	\centering				
	\includegraphics[width=8.5cm]{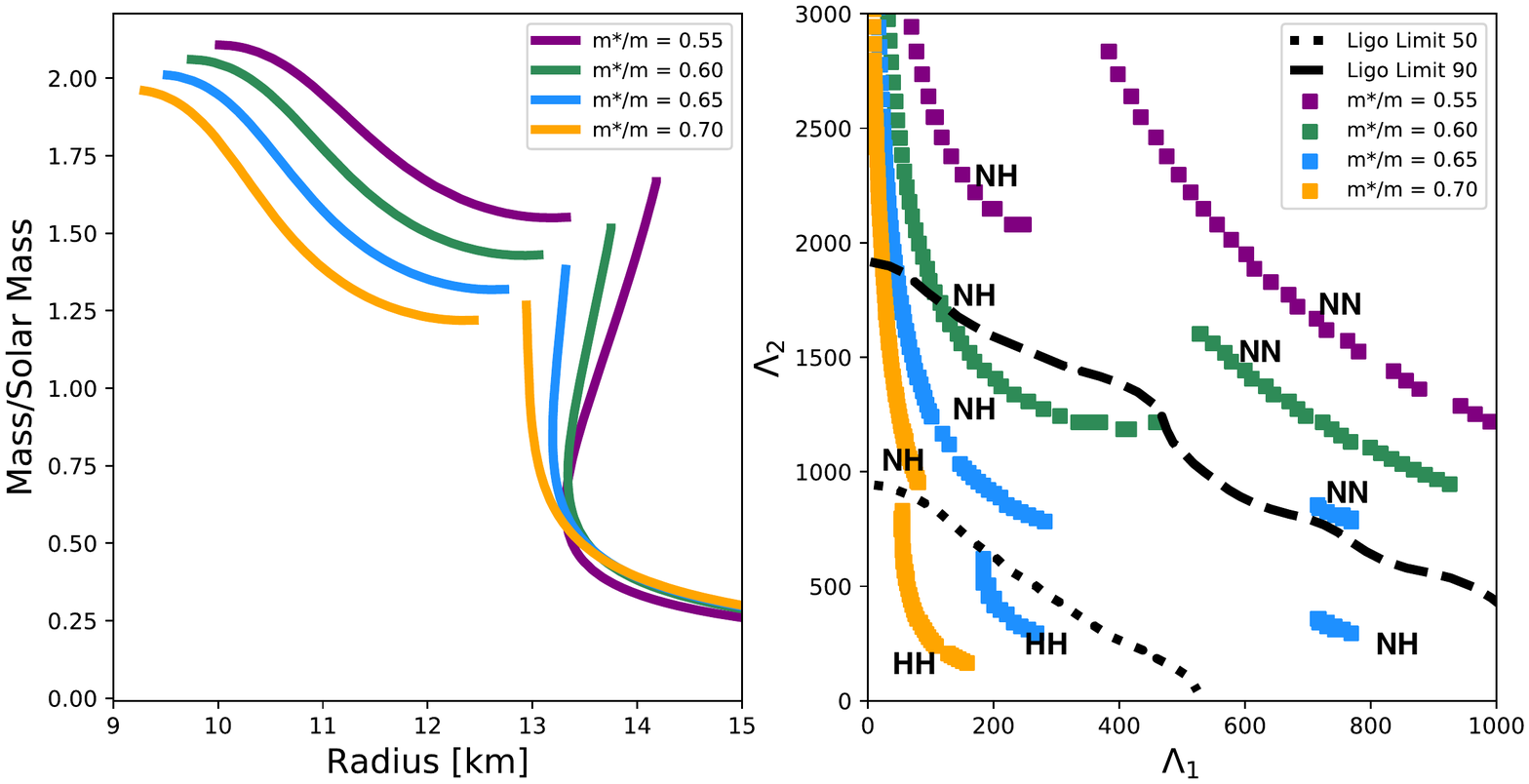}
	\caption{\footnotesize Category \rom{3} phase transitions with the parameters $p_{trans}= 43\,\mathrm{MeV/fm^3}$ and $\Delta\epsilon=350\,\mathrm{MeV/fm^3}$ are depicted. 
	There are fewer NN combinations than in the pure case (see Fig. \ref{LvLCatpure}), since the neutron star branch in the mass-radius relation contains fewer stars. However, the remaining NN combinations do not change their position. The NH combinations and the HH combinations move further into the LIGO credibility limit or closer to it. The case $m^*/m=0.65$ is a special case, where a single EoS exhibits possible NN, NH and HH combinations.}
	\label{LvLCatIII}
\end{figure}

\begin{figure}
	\centering				
	\includegraphics[width=8.5cm]{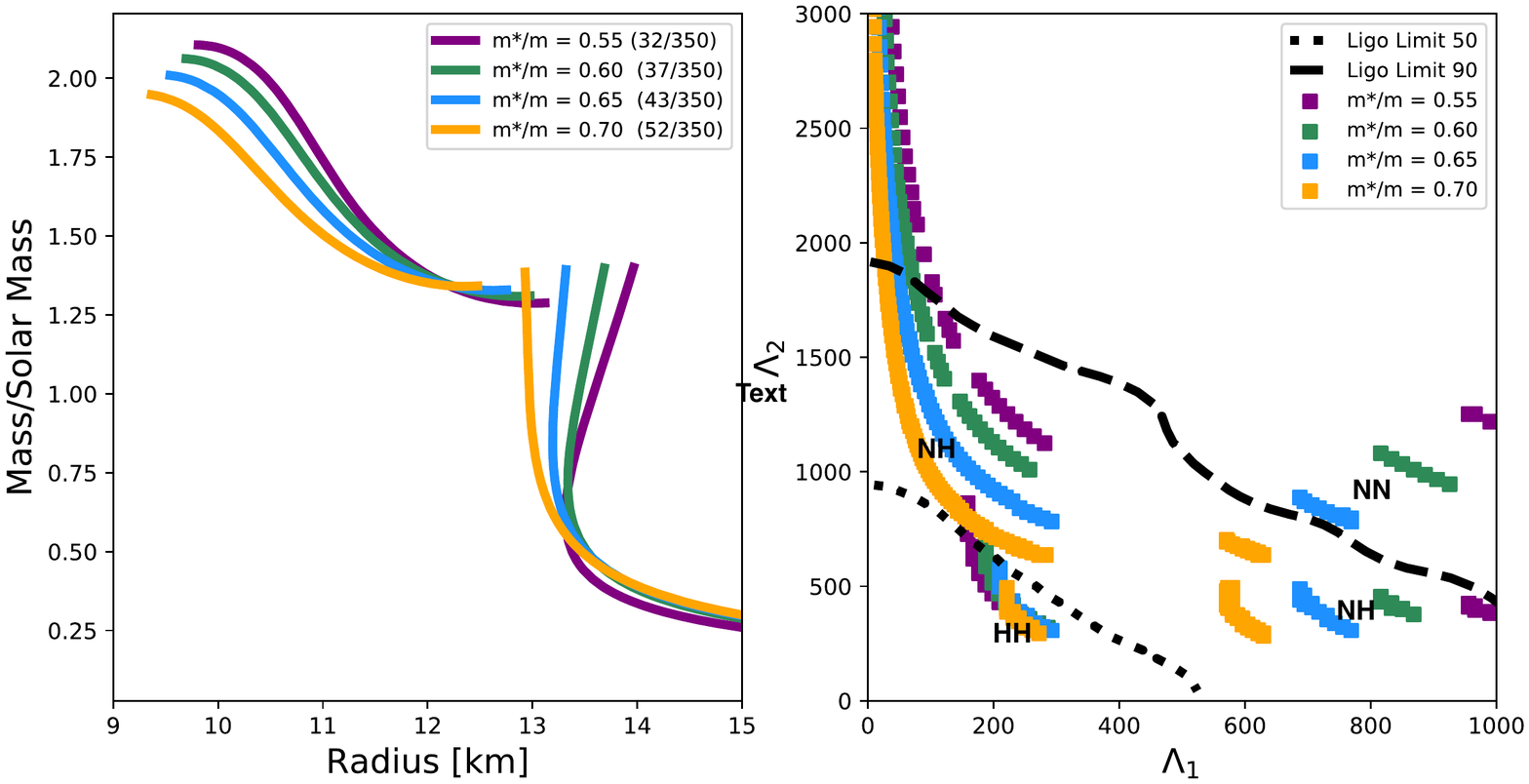}
	\caption{\footnotesize The special case for every considered effective mass is depicted. The phase transition is located at the necessary mass to generate HH, NH and NN combinations for every value of $m^*/m$, which in the case of GW170817, is roughly $1.4M_\odot$. The parameters are written in the legend in the order ($p_{trans}$/$\Delta\epsilon$) in units of $\mathrm{MeV/fm^3}$.}
	\label{LvLCatIIIs}
\end{figure}

The values  $p_{trans}= 43\,\mathrm{MeV/fm^3}$ and $\Delta\epsilon=350\,\mathrm{MeV/fm^3}$ can generate category \rom{3} solutions for all values of $m^*/m$ considered. This is depicted in figure \ref{LvLCatIII}. The stiffest EoS is completely outside of the LIGO credibility level. However, even for the stiffest case the NH pairs are closer to the credibility limit than the pure NN case. The $m^*/m=0.60$ EoS is the first case, where the phase transition improves the compatibility of an EoS with the LIGO measurement, by moving some NH combinations into the 90\% credibility area, where the pure NN case would be outside of it.\\
In our previous publication \cite{Christian:2018jyd} we found a special case for a transition at values of $p_{trans}= 43\,\mathrm{MeV/fm^3}$ and $\Delta\epsilon=350\,\mathrm{MeV/fm^3}$,  where NN, NH and HH combinations were generated by a single EoS. The NH pairs are located in two areas, one above the $\Lambda_{1}=\Lambda_{2}$ limit and one below. The latter case is generated by so called rising twins, where the more massive twin star has a larger radius \cite{Schertler:2000xq}. The hadronic EoS in that case was the DD2 equation by Typel et al. \cite{Typel:2009sy}, which has an effective nucleon mass of $m^*/m=0.6255$. 
A similar special case can be found for the EoS covered in this work, for an effective mass of $m^*/m=0.65$. However, if so desired a special case can be realized for any category \rom{3} EoS, if the transition parameters are chosen accordingly (see figure \ref{LvLCatIIIs}). The NN pairs of the $m^*/m=0.65$ case are already at the border of the credible area and the NH pairs can move even further into it. The HH pairs of the  $m^*/m=0.65$ reach below even the 50\% credibility limit. The $m^*/m=0.70$ case does not exhibit NN combinations, but the NH pairs are located nearly completely in the credibility limit, while the HH pairs are below the 50\% credibility limit. The $m^*/m=0.75$ case is missing, because it is not possible reach the $2M_\odot$ requirement with a CIII phase transition.


\begin{figure}
	\centering				
	\includegraphics[width=8.5cm]{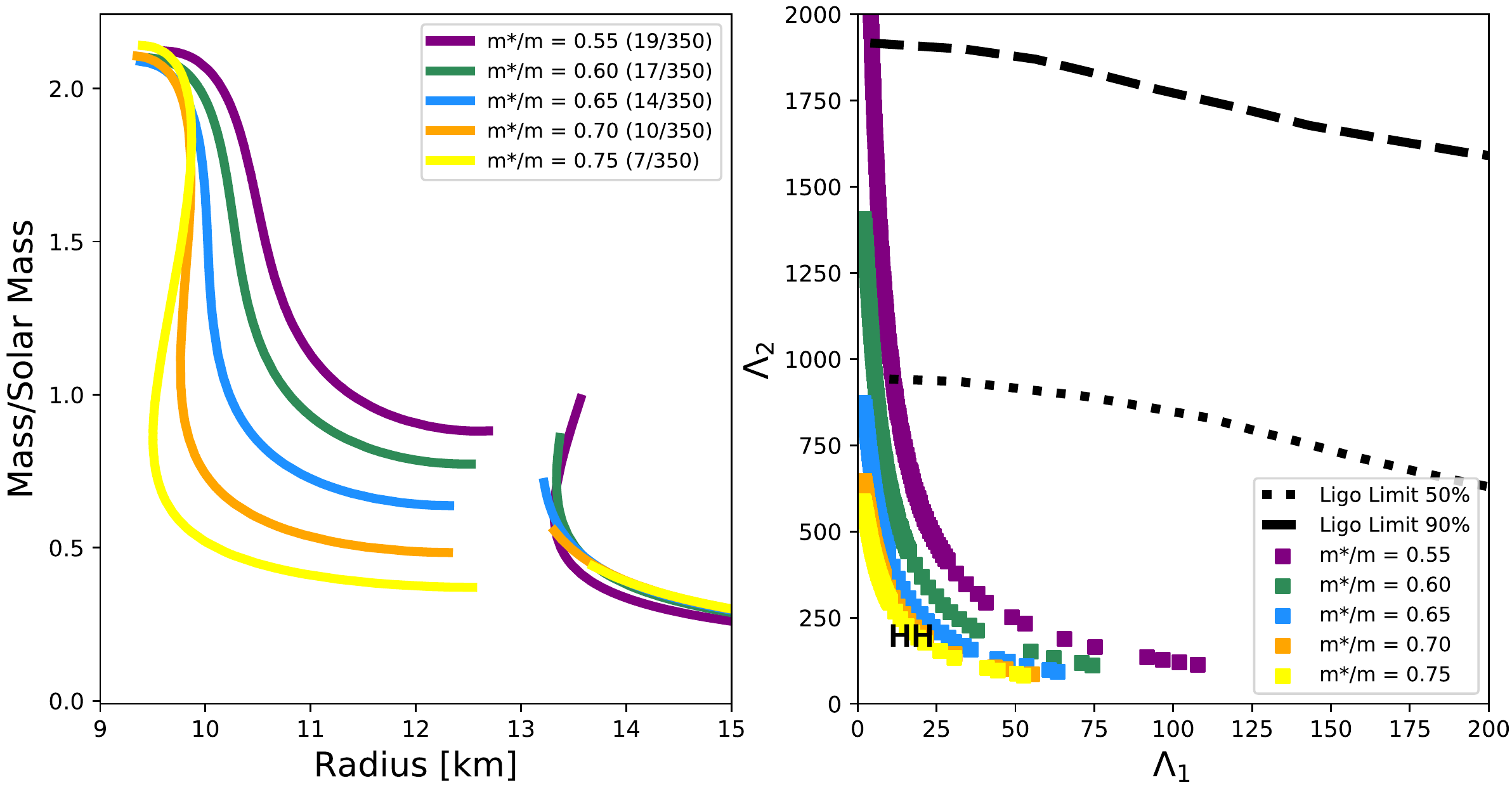}
	\caption{\footnotesize Cases of category IV phase transitions are depicted, the parameters are written behind the corresponding values of $m^*/m$ in the order ($p_{trans}$/$\Delta\epsilon$) in units of $\mathrm{MeV/fm^3}$. Category IV is dominated by the EoS describing quark matter. As a result the second branch is incredibly stiff and the effective mass has virtually no impact on the EoS. Since only hybrid stars can be combined with other hybrid stars to find the possible areas in the $\Lambda_{1}$-$\Lambda_{2}$ plot on the right hand side the possible combinations are very close to each other, even though their mass-radius relations (on the left) appear to be very different.} 
	\label{LvLCatIV}
\end{figure}

In contrast to the previous categories it is not difficult to find an EoS in category \rom{4}, that produces combinations inside the 50\% credibility limit. This is because the early phase transition makes the quark matter equation of state more dominant and this EoS was chosen specifically to be the most stiffest possible equation consistent with causality. 
Due to the identical quark matter EoS in all cases we chose to depict different phase transition parameters in Fig. \ref{LvLCatIV}, as similar values would generate mass-radius relations that are nearly on top of each other. Still the resulting combinations in the the $\Lambda_{1}-\Lambda_{2}$ plot are close together (see the right side of Fig. \ref{LvLCatIV}). By definition it is only possible to find HH lines in a category \rom{4} case. 

In table \ref{Tablenew} the compatibility of the four categories with GW170817 in dependence on the effective nucleon mass $m^*/m$ is broadly summarized. In the table category \rom{1} is written down as I/0, where 0 means "no category". The "x" symbol marks cases, where a phase transition fulfills the $2M_\odot$ constraint, but no combinations of neutron stars are located within the LIGO credibility limit. The "y" symbol marks cases, where any combination is located within the credibility limit. The "o" is used, when the most compact pairs are directly at the credibility limit. 
A phase transition of category \rom{1} does not change the compatibility of any of the hadronic EoS with the GW170817 data, which means that only the cases $0.65 \le m^*/m \le 0.75$ are within the credibility limit, with $m^*/m = 0.65$ at its very border.\\
The same is true for a category \rom{2} phase transitions. However, it is important to stress, that only effective nucleon masses of $m^*/m \le 0.70$ can be realized with a category \rom{1} or \rom{2} phase transition. The $m^*/m=0.75$ case is too soft to generate a stable second branch at the high values of $p_{trans}$ required for the first two categories.\\
A phase transition of category \rom{3} can lead to NH and HH combinations within the LIGO credibility limit for the cases $m^*/m \le 0.70$. This means, that the case $m^*/m = 0.70$ is the only case, that can generate NN and NH pairs that are completely within the credibility limit. The $m^*/m = 0.75$ can not be realized with a phase transition, that generates a stable second branch. However, this configuration can not be considered a category \rom{3} case, as the second branch can not reach $2M_\odot$.\\
It is only possible to find a $m^*/m = 0.75$ case, that generates a second branch and has a maximal mass that exceeds $2M_\odot$ if the first branch has its maximum below $1M_\odot$. This means, that all cases $m^*/m$ can generate a stable second branch in the form of a category \rom{4} phase transition. Category \rom{4} phase transitions generate only HH combinations,these combinations are very compact and as a result all examined cases of $m^*/m$ are within the 50\% credibility limit.\\
However, due to the early phase transition the influence of $m^*/m$ on the mass-radius relation is negligible. As a result no meaningful statement about the influence of the effective nucleon mass on a category \rom{4} phase transition can be made.

\begin{table}
	\begin{center}
		\begin{tabular}{|c|c|c|c|c|c|}
			\hline
			\hspace{.2cm} Category \hspace{.2cm} & \hspace{.01cm} 0.55
			\hspace{.1cm} & \hspace{.2cm} 0.60 \hspace{.2cm} & \hspace{.1cm}
			0.65 \hspace{.1cm} & \hspace{.1cm}
			0.70 \hspace{.1cm} & \hspace{.1cm}
			0.75 \hspace{.1cm}\\
			\cline{1-6}
			I/0 & x & x & o & y & y\\ 
			II & x & x & o & y & n.a.\\ 
			III & y & y & y & y  & n.a.\\ 
			IV & y & y & y & y & y\\
			\hline
		\end{tabular}
		\caption[Table: $m^*/m$ supported by tidal deformability observation]{These are the cases of $m^*/m$ supported by tidal deformability observation. The o denotes the cases where the line is at the 90$\%$ credibility limit, y is below and x is above.}
		\label{Tablenew}
	\end{center}
\end{table}

\section{A Nicer View on twin stars}
\begin{figure}[H]
	\centering				
	\includegraphics[width=8.5cm]{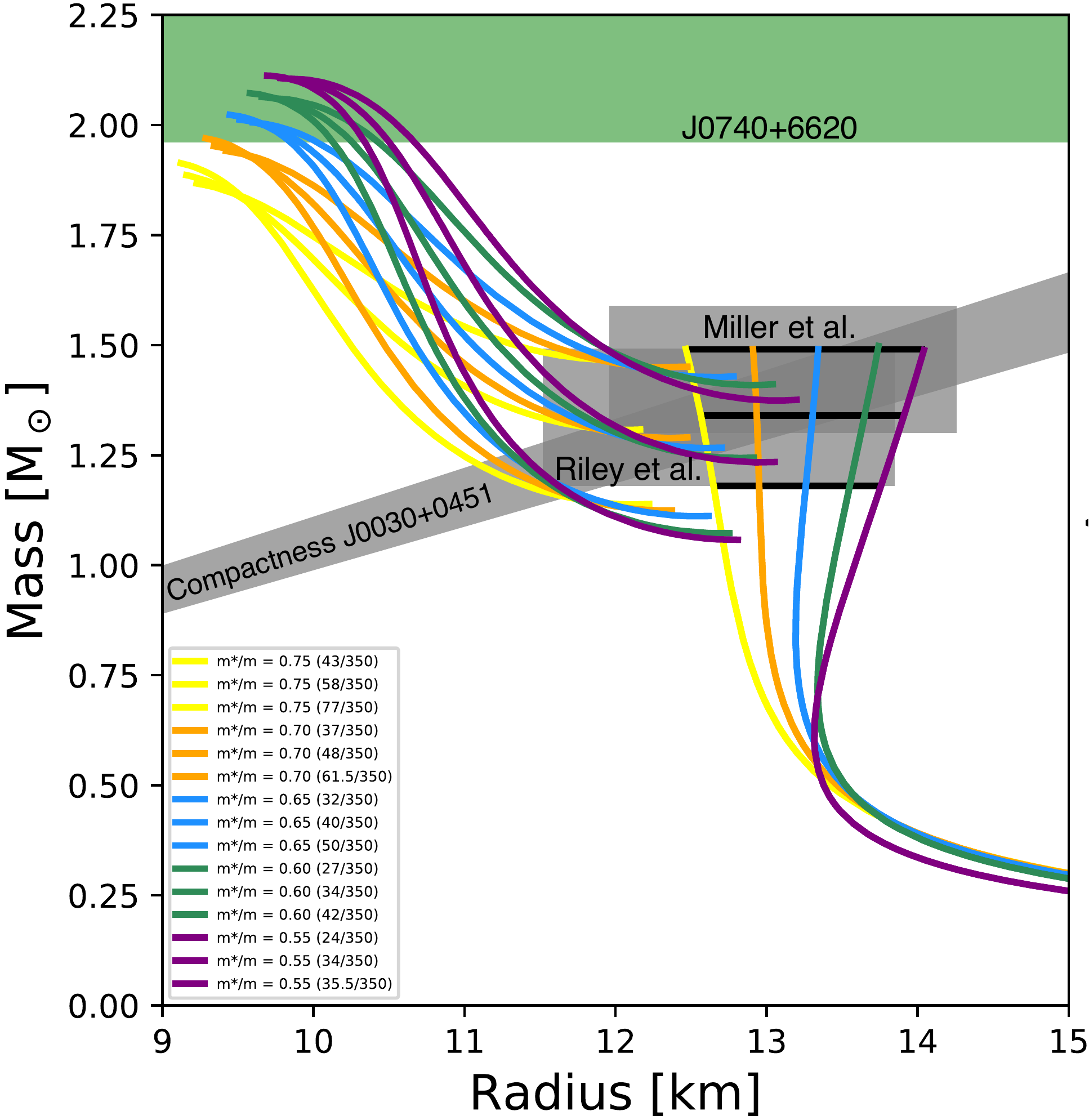}
	\caption{\footnotesize Mass-radius relations for category \rom{3} phase transitions are depicted for all considered effective nucleon masses. The constraints for the J0030+0451 measurement by NICER 
	are taken from refs. \cite{Riley:2019yda,Miller:2019cac} and are shaded gray. The $2M_\odot$ constraint from J0740+6620 \cite{Cromartie:2019kug} is shaded green.  All cases of $m^*/m$ can generate neutron stars and hybrid stars within the NICER likelihood, if the transition parameters are chosen accordingly. The case $m^*/m=0.75$ does not meet the $2M_\odot$ constraint if a phase transition takes place. The black straight lines indicate the maximal mass of the hadronic branch for the respective transitional pressure.} 
	\label{Nicerview}
\end{figure}

The recently released mass and radius measurements of the pulsar J0030+0451 by the NICER program \cite{Riley:2019yda,Miller:2019cac,Raaijmakers:2019qny} can be used to constrain the EoSs discussed previously. NICER measures neutron star radii by observing hotspots on the pulsars surface. Depending on the model used to place these hotspots, two different masses and radii are determined. Riley at al. find a mass of $1.34^{+0.15}_{-0.16}M_\odot$ with a radius of $12.71^{+1.14}_{-1.19}\mathrm{km}$ \cite{Riley:2019yda}, while Miller et al. find $1.44^{+0.15}_{-0.14}M_\odot$ with a radius of $13.02^{+1.24}_{-1.06}\mathrm{km}$ \cite{Miller:2019cac}. The compactness is determined more precisely and in both cases given as: $MG/Rc^2=0.16\pm0.01$.\\
In Fig. \ref{Nicerview} a sample of category \rom{3} EoSs is depicted, with the constraints from NICER shaded gray and the $2M_\odot$ constraint from J0740+6620 \cite{Cromartie:2019kug} shaded green. We find, that for the pure hadronic cases, all considered effective masses generate neutron stars, that fit within the mass-radius range determined by Miller et al. \cite{Miller:2019cac}. The cases $m^*/m \ge 0.55$ are within the range determined by Riley et al. \cite{Riley:2019yda} as well, only the $m^*/m=0.55$ case is outside the range.\\
By definition, only category \rom{2} - \rom{4} phase transitions can support hybrid stars, that fulfill the NICER constraints. Category \rom{1} has to meet the constraints with its hadronic branch. However, category \rom{2} phase transitions generate massive hybrid stars, which usually are at higher masses than the constraint as well. Category \rom{3} phase transitions take place at a mass range, that is within the NICER likelihood. As a result hybrid stars and pure hadronic stars, that fit within the constraint can be found for all effective nucleon masses. The black straight lines in Fig. \ref{Nicerview} indicate the maximum of the hadronic branch. The lowest maximal masses are generated by the lowest transitional pressures of the respective cases. The $m^*/m=0.75$ case can not reach the $2M_\odot$ constraint, as mentioned previously, however the pure hadronic $m^*/m=0.75$ case fits well with the Riley et al. mass and radius data \cite{Riley:2019yda}.\\
Due to the comparatively small uncertainty in radius category \rom{4} phase transitions that generate neutron stars within the constraints from either Riley et al. or Miller et al. are impossible to find. The hadronic branch ends before the minimal mass is reached. The hybrid star branch would be located at smaller radii than required. This behavior can be seen for the earliest phase transitions of the category \rom{3} examples in Fig. \ref{Nicerview} as well. Therefore we can state, that a strong phase transition is only compatible with the NICER constraints if the maximal mass of the hadronic branch is greater than the minimal mass of the NICER measurement. This can be related to the transitional pressure and the density. We find, that strong phase transitions are not viable for  densities below $n \lesssim 1.7\,n_0$. We consider phase transitions  "strong" if $\Delta\epsilon\ge350\mathrm{MeV/fm^3}$. This value is the lowest value for of discontinuity in energy density that generates a visible difference between the hadronic maximum and the hybrid star minimum of  about $0.1M_\odot$ for category \rom{4} cases. We used the explicit radii from Riley et al. \cite{Riley:2019yda} and Miller et al. \cite{Miller:2019cac} instead of the corresponding likelihood ellipses. When considering the $2\sigma$ likelihood ellipses \cite{Raaijmakers:2019qny} the constraints become weaker. 
However, a phase transition with parameters $n\lesssim1.4n_0$ and $\Delta\epsilon\ge350\mathrm{MeV/fm^3}$ is still outside the $2\sigma$ likelihood constraints from the NICER measurement.

\section{Conclusion}
By employing a parameterized relativistic mean field equation of state we explored how the stiffness of a hadronic EoS influences the tidal deformability of an EoS featuring a phase transition from hadronic to quark matter. This phase transitions are chosen to represent the four categories of twin stars \cite{Christian:2017jni}. As stated in our previous work \cite{Christian:2018jyd} an EoSs with a phase transition can generate hybrid-hybrid, neutron-hybrid and neutron-neutron combinations, depending on the location of the phase transition in the mass-radius relation. We consider effective nucleon masses from $m^*/m=0.75$ to $m^*/m=0.55$, where a larger effective mass corresponds to a softer EoS. Like Hornick et al. \cite{Hornick:2018kfi}, we find, that only the pure hadronic cases for $m^*/m\ge0.65$ are compatible with the GW170817 data \cite{Abbott:2018wiz}. The $m^*/m=0.75$ case is to soft to generate a stable second branch, that fulfills the requirement for a category \rom{1},\rom{2} or \rom{3} phase transition. Future measurement of neutron stars with masses above $2M_\odot$ might cause similar problems for the $m^*/m=0.70$ case. At the time of this work the maximal masses of the $m^*/m=0.70$ category \rom{2} and \rom{3} phase transitions are still within the margin of error of the most massive know neutron stars \cite{Demorest:2010bx,Antoniadis:2013pzd,Fonseca:2016tux,Cromartie:2019kug}.\\
Due to the higher compactness of hybrid stars, the cases $m^*/m\le 0.70$ can generate neutron star pairs, from a category \rom{3} phase transtition, deeper within the credibility limit than they could without a phase transition. This means, that even the cases $m^*/m=0.55$ and $m^*/m=0.60$, which are on their own too stiff to allow for pairs of neutron stars with sufficiently small values of tidal deformability, can generate combinations within the credibility limit. The benefits of a phase transition in regards to an EoS's compatibility with the LIGO data have been shown before \cite{Paschalidis:2017qmb,Alvarez-Castillo:2018pve,Christian:2018jyd,Montana:2018bkb}.\\
Only category \rom{4} can be realized for all examined values of $m^*/m$, this is because the resulting EoSs can be considered independent form $m^*/m$ due to the early phase transition. However, the recent results form NICER \cite{Raaijmakers:2019qny} are incompatible with a category \rom{4} phase transition and furthermore exclude strong phase transitions at densities of $n \lesssim 1.7\,n_0$, where a visible jump in mass of  $\Delta M\ge0.1M_\odot$ at the point of transition occurs.

\begin{acknowledgments}
The authors thank Andreas Zacchi for helpful discussions. JS acknowledges support from the Helmholtz International Center for FAIR (HIC for FAIR). JEC is a recipient of the Carlo and Karin Giersch Scholarship of the Giersch foundation. 
\end{acknowledgments}
\bibliographystyle{apsrev4-1}
\bibliography{neue_bib_JSB}

\begin{thebibliography}{59}%
\makeatletter
\providecommand \@ifxundefined [1]{%
 \@ifx{#1\undefined}
}%
\providecommand \@ifnum [1]{%
 \ifnum #1\expandafter \@firstoftwo
 \else \expandafter \@secondoftwo
 \fi
}%
\providecommand \@ifx [1]{%
 \ifx #1\expandafter \@firstoftwo
 \else \expandafter \@secondoftwo
 \fi
}%
\providecommand \natexlab [1]{#1}%
\providecommand \enquote  [1]{``#1''}%
\providecommand \bibnamefont  [1]{#1}%
\providecommand \bibfnamefont [1]{#1}%
\providecommand \citenamefont [1]{#1}%
\providecommand \href@noop [0]{\@secondoftwo}%
\providecommand \href [0]{\begingroup \@sanitize@url \@href}%
\providecommand \@href[1]{\@@startlink{#1}\@@href}%
\providecommand \@@href[1]{\endgroup#1\@@endlink}%
\providecommand \@sanitize@url [0]{\catcode `\\12\catcode `\$12\catcode
  `\&12\catcode `\#12\catcode `\^12\catcode `\_12\catcode `\%12\relax}%
\providecommand \@@startlink[1]{}%
\providecommand \@@endlink[0]{}%
\providecommand \url  [0]{\begingroup\@sanitize@url \@url }%
\providecommand \@url [1]{\endgroup\@href {#1}{\urlprefix }}%
\providecommand \urlprefix  [0]{URL }%
\providecommand \Eprint [0]{\href }%
\providecommand \doibase [0]{http://dx.doi.org/}%
\providecommand \selectlanguage [0]{\@gobble}%
\providecommand \bibinfo  [0]{\@secondoftwo}%
\providecommand \bibfield  [0]{\@secondoftwo}%
\providecommand \translation [1]{[#1]}%
\providecommand \BibitemOpen [0]{}%
\providecommand \bibitemStop [0]{}%
\providecommand \bibitemNoStop [0]{.\EOS\space}%
\providecommand \EOS [0]{\spacefactor3000\relax}%
\providecommand \BibitemShut  [1]{\csname bibitem#1\endcsname}%
\let\auto@bib@innerbib\@empty
\bibitem [{\citenamefont {Ivanenko}\ and\ \citenamefont
  {Kurdgelaidze}(1965)}]{Ivanenko:1965dg}%
  \BibitemOpen
  \bibfield  {author} {\bibinfo {author} {\bibfnamefont {D.~D.}\ \bibnamefont
  {Ivanenko}}\ and\ \bibinfo {author} {\bibfnamefont {D.~F.}\ \bibnamefont
  {Kurdgelaidze}},\ }\href@noop {} {\bibfield  {journal} {\bibinfo  {journal}
  {Astrophys.}\ }\textbf {\bibinfo {volume} {1}},\ \bibinfo {pages} {251}
  (\bibinfo {year} {1965})}\BibitemShut {NoStop}%
\bibitem [{\citenamefont {Itoh}(1970)}]{Itoh:1970uw}%
  \BibitemOpen
  \bibfield  {author} {\bibinfo {author} {\bibfnamefont {N.}~\bibnamefont
  {Itoh}},\ }\href {\doibase 10.1143/PTP.44.291} {\bibfield  {journal}
  {\bibinfo  {journal} {Prog.Theor.Phys.}\ }\textbf {\bibinfo {volume} {44}},\
  \bibinfo {pages} {291} (\bibinfo {year} {1970})}\BibitemShut {NoStop}%
\bibitem [{\citenamefont {Alford}\ \emph {et~al.}(2005)\citenamefont {Alford},
  \citenamefont {Braby}, \citenamefont {Paris},\ and\ \citenamefont
  {Reddy}}]{Alford:2004pf}%
  \BibitemOpen
  \bibfield  {author} {\bibinfo {author} {\bibfnamefont {M.}~\bibnamefont
  {Alford}}, \bibinfo {author} {\bibfnamefont {M.}~\bibnamefont {Braby}},
  \bibinfo {author} {\bibfnamefont {M.}~\bibnamefont {Paris}}, \ and\ \bibinfo
  {author} {\bibfnamefont {S.}~\bibnamefont {Reddy}},\ }\href {\doibase
  10.1086/430902} {\bibfield  {journal} {\bibinfo  {journal} {Astrophys.J.}\
  }\textbf {\bibinfo {volume} {629}},\ \bibinfo {pages} {969} (\bibinfo {year}
  {2005})},\ \Eprint {http://arxiv.org/abs/nucl-th/0411016}
  {arXiv:nucl-th/0411016 [nucl-th]} \BibitemShut {NoStop}%
\bibitem [{\citenamefont {Coelho}\ \emph {et~al.}(2010)\citenamefont {Coelho},
  \citenamefont {Lenzi}, \citenamefont {Malheiro}, \citenamefont {Marinho},\
  and\ \citenamefont {Fiolhais}}]{Coelho:2010fv}%
  \BibitemOpen
  \bibfield  {author} {\bibinfo {author} {\bibfnamefont {J.}~\bibnamefont
  {Coelho}}, \bibinfo {author} {\bibfnamefont {C.}~\bibnamefont {Lenzi}},
  \bibinfo {author} {\bibfnamefont {M.}~\bibnamefont {Malheiro}}, \bibinfo
  {author} {\bibfnamefont {J.}~\bibnamefont {Marinho}, \bibfnamefont {R.M.}}, \
  and\ \bibinfo {author} {\bibfnamefont {M.}~\bibnamefont {Fiolhais}},\ }\href
  {\doibase 10.1142/S0218271810017597} {\bibfield  {journal} {\bibinfo
  {journal} {Int.J.Mod.Phys.}\ }\textbf {\bibinfo {volume} {D19}},\ \bibinfo
  {pages} {1521} (\bibinfo {year} {2010})},\ \Eprint
  {http://arxiv.org/abs/1001.1661} {arXiv:1001.1661 [nucl-th]} \BibitemShut
  {NoStop}%
\bibitem [{\citenamefont {Chen}\ \emph {et~al.}(2011)\citenamefont {Chen},
  \citenamefont {Baldo}, \citenamefont {Burgio},\ and\ \citenamefont
  {Schulze}}]{Chen:2011my}%
  \BibitemOpen
  \bibfield  {author} {\bibinfo {author} {\bibfnamefont {H.}~\bibnamefont
  {Chen}}, \bibinfo {author} {\bibfnamefont {M.}~\bibnamefont {Baldo}},
  \bibinfo {author} {\bibfnamefont {G.}~\bibnamefont {Burgio}}, \ and\ \bibinfo
  {author} {\bibfnamefont {H.-J.}\ \bibnamefont {Schulze}},\ }\href {\doibase
  10.1103/PhysRevD.84.105023} {\bibfield  {journal} {\bibinfo  {journal}
  {Phys.Rev.}\ }\textbf {\bibinfo {volume} {D84}},\ \bibinfo {pages} {105023}
  (\bibinfo {year} {2011})},\ \Eprint {http://arxiv.org/abs/1107.2497}
  {arXiv:1107.2497 [nucl-th]} \BibitemShut {NoStop}%
\bibitem [{\citenamefont {Masuda}\ \emph {et~al.}(2013)\citenamefont {Masuda},
  \citenamefont {Hatsuda},\ and\ \citenamefont {Takatsuka}}]{Masuda:2012kf}%
  \BibitemOpen
  \bibfield  {author} {\bibinfo {author} {\bibfnamefont {K.}~\bibnamefont
  {Masuda}}, \bibinfo {author} {\bibfnamefont {T.}~\bibnamefont {Hatsuda}}, \
  and\ \bibinfo {author} {\bibfnamefont {T.}~\bibnamefont {Takatsuka}},\ }\href
  {\doibase 10.1088/0004-637X/764/1/12} {\bibfield  {journal} {\bibinfo
  {journal} {Astrophys. J.}\ }\textbf {\bibinfo {volume} {764}},\ \bibinfo
  {pages} {12} (\bibinfo {year} {2013})},\ \Eprint
  {http://arxiv.org/abs/1205.3621} {arXiv:1205.3621 [nucl-th]} \BibitemShut
  {NoStop}%
\bibitem [{\citenamefont {Yasutake}\ \emph {et~al.}(2014)\citenamefont
  {Yasutake}, \citenamefont {Lastowiecki}, \citenamefont {Benic}, \citenamefont
  {Blaschke}, \citenamefont {Maruyama} \emph {et~al.}}]{Yasutake:2014oxa}%
  \BibitemOpen
  \bibfield  {author} {\bibinfo {author} {\bibfnamefont {N.}~\bibnamefont
  {Yasutake}}, \bibinfo {author} {\bibfnamefont {R.}~\bibnamefont
  {Lastowiecki}}, \bibinfo {author} {\bibfnamefont {S.}~\bibnamefont {Benic}},
  \bibinfo {author} {\bibfnamefont {D.}~\bibnamefont {Blaschke}}, \bibinfo
  {author} {\bibfnamefont {T.}~\bibnamefont {Maruyama}},  \emph {et~al.},\
  }\href {\doibase 10.1103/PhysRevC.89.065803} {\bibfield  {journal} {\bibinfo
  {journal} {Phys.Rev.}\ }\textbf {\bibinfo {volume} {C89}},\ \bibinfo {pages}
  {065803} (\bibinfo {year} {2014})},\ \Eprint {http://arxiv.org/abs/1403.7492}
  {arXiv:1403.7492 [astro-ph.HE]} \BibitemShut {NoStop}%
\bibitem [{\citenamefont {Zacchi}\ \emph {et~al.}(2016)\citenamefont {Zacchi},
  \citenamefont {Hanauske},\ and\ \citenamefont
  {Schaffner-Bielich}}]{Zacchi:2015oma}%
  \BibitemOpen
  \bibfield  {author} {\bibinfo {author} {\bibfnamefont {A.}~\bibnamefont
  {Zacchi}}, \bibinfo {author} {\bibfnamefont {M.}~\bibnamefont {Hanauske}}, \
  and\ \bibinfo {author} {\bibfnamefont {J.}~\bibnamefont
  {Schaffner-Bielich}},\ }\href {\doibase 10.1103/PhysRevD.93.065011}
  {\bibfield  {journal} {\bibinfo  {journal} {Phys. Rev.}\ }\textbf {\bibinfo
  {volume} {D93}},\ \bibinfo {pages} {065011} (\bibinfo {year} {2016})},\
  \Eprint {http://arxiv.org/abs/1510.00180} {arXiv:1510.00180 [nucl-th]}
  \BibitemShut {NoStop}%
\bibitem [{\citenamefont {K\"ampfer}(1981)}]{Kampfer:1981yr}%
  \BibitemOpen
  \bibfield  {author} {\bibinfo {author} {\bibfnamefont {B.}~\bibnamefont
  {K\"ampfer}},\ }\href {\doibase 10.1088/0305-4470/14/11/009} {\bibfield
  {journal} {\bibinfo  {journal} {J.Phys.}\ }\textbf {\bibinfo {volume}
  {A14}},\ \bibinfo {pages} {L471} (\bibinfo {year} {1981})}\BibitemShut
  {NoStop}%
\bibitem [{\citenamefont {Glendenning}\ and\ \citenamefont
  {Kettner}(2000)}]{Glendenning:1998ag}%
  \BibitemOpen
  \bibfield  {author} {\bibinfo {author} {\bibfnamefont {N.~K.}\ \bibnamefont
  {Glendenning}}\ and\ \bibinfo {author} {\bibfnamefont {C.}~\bibnamefont
  {Kettner}},\ }\href@noop {} {\bibfield  {journal} {\bibinfo  {journal}
  {Astron. Astrophys.}\ }\textbf {\bibinfo {volume} {353}},\ \bibinfo {pages}
  {L9} (\bibinfo {year} {2000})},\ \Eprint
  {http://arxiv.org/abs/astro-ph/9807155} {astro-ph/9807155} \BibitemShut
  {NoStop}%
\bibitem [{\citenamefont {Schertler}\ \emph {et~al.}(2000)\citenamefont
  {Schertler}, \citenamefont {Greiner}, \citenamefont {Schaffner-Bielich},\
  and\ \citenamefont {Thoma}}]{Schertler:2000xq}%
  \BibitemOpen
  \bibfield  {author} {\bibinfo {author} {\bibfnamefont {K.}~\bibnamefont
  {Schertler}}, \bibinfo {author} {\bibfnamefont {C.}~\bibnamefont {Greiner}},
  \bibinfo {author} {\bibfnamefont {J.}~\bibnamefont {Schaffner-Bielich}}, \
  and\ \bibinfo {author} {\bibfnamefont {M.~H.}\ \bibnamefont {Thoma}},\
  }\href@noop {} {\bibfield  {journal} {\bibinfo  {journal} {Nucl. Phys.}\
  }\textbf {\bibinfo {volume} {A677}},\ \bibinfo {pages} {463} (\bibinfo {year}
  {2000})},\ \Eprint {http://arxiv.org/abs/astro-ph/0001467} {astro-ph/0001467}
  \BibitemShut {NoStop}%
\bibitem [{\citenamefont {Schaffner-Bielich}\ \emph {et~al.}(2002)\citenamefont
  {Schaffner-Bielich}, \citenamefont {Hanauske}, \citenamefont {St{\"o}cker},\
  and\ \citenamefont {Greiner}}]{SchaffnerBielich:2002ki}%
  \BibitemOpen
  \bibfield  {author} {\bibinfo {author} {\bibfnamefont {J.}~\bibnamefont
  {Schaffner-Bielich}}, \bibinfo {author} {\bibfnamefont {M.}~\bibnamefont
  {Hanauske}}, \bibinfo {author} {\bibfnamefont {H.}~\bibnamefont
  {St{\"o}cker}}, \ and\ \bibinfo {author} {\bibfnamefont {W.}~\bibnamefont
  {Greiner}},\ }\href@noop {} {\bibfield  {journal} {\bibinfo  {journal} {Phys.
  Rev. Lett.}\ }\textbf {\bibinfo {volume} {89}},\ \bibinfo {pages} {171101}
  (\bibinfo {year} {2002})},\ \Eprint {http://arxiv.org/abs/astro-ph/0005490}
  {astro-ph/0005490} \BibitemShut {NoStop}%
\bibitem [{\citenamefont {Zdunik}\ and\ \citenamefont
  {Haensel}(2013)}]{Zdunik:2012dj}%
  \BibitemOpen
  \bibfield  {author} {\bibinfo {author} {\bibfnamefont {J.}~\bibnamefont
  {Zdunik}}\ and\ \bibinfo {author} {\bibfnamefont {P.}~\bibnamefont
  {Haensel}},\ }\href {\doibase 10.1051/0004-6361/201220697} {\bibfield
  {journal} {\bibinfo  {journal} {Astron.Astrophys.}\ }\textbf {\bibinfo
  {volume} {551}},\ \bibinfo {pages} {A61} (\bibinfo {year} {2013})},\ \Eprint
  {http://arxiv.org/abs/1211.1231} {arXiv:1211.1231 [astro-ph.SR]} \BibitemShut
  {NoStop}%
\bibitem [{\citenamefont {Alford}\ \emph {et~al.}(2015)\citenamefont {Alford},
  \citenamefont {Burgio}, \citenamefont {Han}, \citenamefont {Taranto},\ and\
  \citenamefont {Zappalà}}]{Alford:2015dpa}%
  \BibitemOpen
  \bibfield  {author} {\bibinfo {author} {\bibfnamefont {M.~G.}\ \bibnamefont
  {Alford}}, \bibinfo {author} {\bibfnamefont {G.~F.}\ \bibnamefont {Burgio}},
  \bibinfo {author} {\bibfnamefont {S.}~\bibnamefont {Han}}, \bibinfo {author}
  {\bibfnamefont {G.}~\bibnamefont {Taranto}}, \ and\ \bibinfo {author}
  {\bibfnamefont {D.}~\bibnamefont {Zappalà}},\ }\href {\doibase
  10.1103/PhysRevD.92.083002} {\bibfield  {journal} {\bibinfo  {journal} {Phys.
  Rev.}\ }\textbf {\bibinfo {volume} {D92}},\ \bibinfo {pages} {083002}
  (\bibinfo {year} {2015})},\ \Eprint {http://arxiv.org/abs/1501.07902}
  {arXiv:1501.07902 [nucl-th]} \BibitemShut {NoStop}%
\bibitem [{\citenamefont {Blaschke}\ and\ \citenamefont
  {Alvarez-Castillo}(2016)}]{Blaschke:2015uva}%
  \BibitemOpen
  \bibfield  {author} {\bibinfo {author} {\bibfnamefont {D.}~\bibnamefont
  {Blaschke}}\ and\ \bibinfo {author} {\bibfnamefont {D.~E.}\ \bibnamefont
  {Alvarez-Castillo}},\ }\bibfield  {booktitle} {\emph {\bibinfo {booktitle}
  {{Proceedings, 11th Conference on Quark Confinement and the Hadron Spectrum
  (Confinement XI): St. Petersburg, Russia, September 8-12, 2014}}},\ }\href
  {\doibase 10.1063/1.4938602} {\bibfield  {journal} {\bibinfo  {journal} {AIP
  Conf. Proc.}\ }\textbf {\bibinfo {volume} {1701}},\ \bibinfo {pages} {020013}
  (\bibinfo {year} {2016})},\ \Eprint {http://arxiv.org/abs/1503.03834}
  {arXiv:1503.03834 [astro-ph.HE]} \BibitemShut {NoStop}%
\bibitem [{\citenamefont {Zacchi}\ \emph {et~al.}(2017)\citenamefont {Zacchi},
  \citenamefont {Tolos},\ and\ \citenamefont
  {Schaffner-Bielich}}]{Zacchi:2016tjw}%
  \BibitemOpen
  \bibfield  {author} {\bibinfo {author} {\bibfnamefont {A.}~\bibnamefont
  {Zacchi}}, \bibinfo {author} {\bibfnamefont {L.}~\bibnamefont {Tolos}}, \
  and\ \bibinfo {author} {\bibfnamefont {J.}~\bibnamefont
  {Schaffner-Bielich}},\ }\href {\doibase 10.1103/PhysRevD.95.103008}
  {\bibfield  {journal} {\bibinfo  {journal} {Phys. Rev.}\ }\textbf {\bibinfo
  {volume} {D95}},\ \bibinfo {pages} {103008} (\bibinfo {year} {2017})},\
  \Eprint {http://arxiv.org/abs/1612.06167} {arXiv:1612.06167 [astro-ph.HE]}
  \BibitemShut {NoStop}%
\bibitem [{\citenamefont {Alford}\ and\ \citenamefont
  {Sedrakian}(2017)}]{Alford:2017qgh}%
  \BibitemOpen
  \bibfield  {author} {\bibinfo {author} {\bibfnamefont {M.~G.}\ \bibnamefont
  {Alford}}\ and\ \bibinfo {author} {\bibfnamefont {A.}~\bibnamefont
  {Sedrakian}},\ }\href {\doibase 10.1103/PhysRevLett.119.161104} {\bibfield
  {journal} {\bibinfo  {journal} {Phys. Rev. Lett.}\ }\textbf {\bibinfo
  {volume} {119}},\ \bibinfo {pages} {161104} (\bibinfo {year} {2017})},\
  \Eprint {http://arxiv.org/abs/1706.01592} {arXiv:1706.01592 [astro-ph.HE]}
  \BibitemShut {NoStop}%
\bibitem [{\citenamefont {Christian}\ \emph {et~al.}(2018)\citenamefont
  {Christian}, \citenamefont {Zacchi},\ and\ \citenamefont
  {Schaffner-Bielich}}]{Christian:2017jni}%
  \BibitemOpen
  \bibfield  {author} {\bibinfo {author} {\bibfnamefont {J.-E.}\ \bibnamefont
  {Christian}}, \bibinfo {author} {\bibfnamefont {A.}~\bibnamefont {Zacchi}}, \
  and\ \bibinfo {author} {\bibfnamefont {J.}~\bibnamefont
  {Schaffner-Bielich}},\ }\href {\doibase 10.1140/epja/i2018-12472-y}
  {\bibfield  {journal} {\bibinfo  {journal} {Eur. Phys. J.}\ }\textbf
  {\bibinfo {volume} {A54}},\ \bibinfo {pages} {28} (\bibinfo {year} {2018})},\
  \Eprint {http://arxiv.org/abs/1707.07524} {arXiv:1707.07524 [astro-ph.HE]}
  \BibitemShut {NoStop}%
\bibitem [{\citenamefont {Blaschke}\ \emph {et~al.}(2020)\citenamefont
  {Blaschke}, \citenamefont {Alvarez-Castillo}, \citenamefont {Ayriyan},
  \citenamefont {Grigorian}, \citenamefont {Lagarni},\ and\ \citenamefont
  {Weber}}]{Blaschke:2019tbh}%
  \BibitemOpen
  \bibfield  {author} {\bibinfo {author} {\bibfnamefont {D.}~\bibnamefont
  {Blaschke}}, \bibinfo {author} {\bibfnamefont {D.~E.}\ \bibnamefont
  {Alvarez-Castillo}}, \bibinfo {author} {\bibfnamefont {A.}~\bibnamefont
  {Ayriyan}}, \bibinfo {author} {\bibfnamefont {H.}~\bibnamefont {Grigorian}},
  \bibinfo {author} {\bibfnamefont {N.~K.}\ \bibnamefont {Lagarni}}, \ and\
  \bibinfo {author} {\bibfnamefont {F.}~\bibnamefont {Weber}}\ }(\bibinfo
  {year} {2020})\ pp.\ \bibinfo {pages} {207--256},\ \Eprint
  {http://arxiv.org/abs/1906.02522} {arXiv:1906.02522 [astro-ph.HE]}
  \BibitemShut {NoStop}%
\bibitem [{\citenamefont {Bodmer}(1971)}]{Bodmer:1971we}%
  \BibitemOpen
  \bibfield  {author} {\bibinfo {author} {\bibfnamefont {A.~R.}\ \bibnamefont
  {Bodmer}},\ }\href@noop {} {\bibfield  {journal} {\bibinfo  {journal} {Phys.
  Rev. D}\ }\textbf {\bibinfo {volume} {4}},\ \bibinfo {pages} {1601} (\bibinfo
  {year} {1971})}\BibitemShut {NoStop}%
\bibitem [{\citenamefont {Haensel}\ \emph {et~al.}(1986)\citenamefont
  {Haensel}, \citenamefont {Zdunik},\ and\ \citenamefont
  {Schaeffer}}]{Haensel:1986qb}%
  \BibitemOpen
  \bibfield  {author} {\bibinfo {author} {\bibfnamefont {P.}~\bibnamefont
  {Haensel}}, \bibinfo {author} {\bibfnamefont {J.~L.}\ \bibnamefont {Zdunik}},
  \ and\ \bibinfo {author} {\bibfnamefont {R.}~\bibnamefont {Schaeffer}},\
  }\href@noop {} {\bibfield  {journal} {\bibinfo  {journal} {Astron.
  Astrophys.}\ }\textbf {\bibinfo {volume} {160}},\ \bibinfo {pages} {121}
  (\bibinfo {year} {1986})}\BibitemShut {NoStop}%
\bibitem [{\citenamefont {Alcock}\ \emph {et~al.}(1986)\citenamefont {Alcock},
  \citenamefont {Farhi},\ and\ \citenamefont {Olinto}}]{Alcock:1986hz}%
  \BibitemOpen
  \bibfield  {author} {\bibinfo {author} {\bibfnamefont {C.}~\bibnamefont
  {Alcock}}, \bibinfo {author} {\bibfnamefont {E.}~\bibnamefont {Farhi}}, \
  and\ \bibinfo {author} {\bibfnamefont {A.}~\bibnamefont {Olinto}},\
  }\href@noop {} {\bibfield  {journal} {\bibinfo  {journal} {Astrophys. J.}\
  }\textbf {\bibinfo {volume} {310}},\ \bibinfo {pages} {261} (\bibinfo {year}
  {1986})}\BibitemShut {NoStop}%
\bibitem [{\citenamefont {Fraga}\ \emph {et~al.}(2002)\citenamefont {Fraga},
  \citenamefont {Pisarski},\ and\ \citenamefont
  {Schaffner-Bielich}}]{Fraga:2001xc}%
  \BibitemOpen
  \bibfield  {author} {\bibinfo {author} {\bibfnamefont {E.~S.}\ \bibnamefont
  {Fraga}}, \bibinfo {author} {\bibfnamefont {R.~D.}\ \bibnamefont {Pisarski}},
  \ and\ \bibinfo {author} {\bibfnamefont {J.}~\bibnamefont
  {Schaffner-Bielich}},\ }\href@noop {} {\bibfield  {journal} {\bibinfo
  {journal} {Nucl. Phys.}\ }\textbf {\bibinfo {volume} {A702}},\ \bibinfo
  {pages} {217} (\bibinfo {year} {2002})},\ \Eprint
  {http://arxiv.org/abs/nucl-th/0110077} {nucl-th/0110077} \BibitemShut
  {NoStop}%
\bibitem [{\citenamefont {Zacchi}\ \emph {et~al.}(2015)\citenamefont {Zacchi},
  \citenamefont {Stiele},\ and\ \citenamefont
  {Schaffner-Bielich}}]{Zacchi:2015lwa}%
  \BibitemOpen
  \bibfield  {author} {\bibinfo {author} {\bibfnamefont {A.}~\bibnamefont
  {Zacchi}}, \bibinfo {author} {\bibfnamefont {R.}~\bibnamefont {Stiele}}, \
  and\ \bibinfo {author} {\bibfnamefont {J.}~\bibnamefont
  {Schaffner-Bielich}},\ }\href {\doibase 10.1103/PhysRevD.92.045022}
  {\bibfield  {journal} {\bibinfo  {journal} {Phys. Rev.}\ }\textbf {\bibinfo
  {volume} {D92}},\ \bibinfo {pages} {045022} (\bibinfo {year} {2015})},\
  \Eprint {http://arxiv.org/abs/1506.01868} {arXiv:1506.01868 [astro-ph.HE]}
  \BibitemShut {NoStop}%
\bibitem [{\citenamefont {Johnson}\ and\ \citenamefont
  {Teller}(1955)}]{PhysRev.98.783}%
  \BibitemOpen
  \bibfield  {author} {\bibinfo {author} {\bibfnamefont {M.~H.}\ \bibnamefont
  {Johnson}}\ and\ \bibinfo {author} {\bibfnamefont {E.}~\bibnamefont
  {Teller}},\ }\href {\doibase 10.1103/PhysRev.98.783} {\bibfield  {journal}
  {\bibinfo  {journal} {Phys. Rev.}\ }\textbf {\bibinfo {volume} {98}},\
  \bibinfo {pages} {783} (\bibinfo {year} {1955})}\BibitemShut {NoStop}%
\bibitem [{\citenamefont {Duerr}(1956)}]{Duerr56}%
  \BibitemOpen
  \bibfield  {author} {\bibinfo {author} {\bibfnamefont {H.-P.}\ \bibnamefont
  {Duerr}},\ }\href@noop {} {\bibfield  {journal} {\bibinfo  {journal} {Phys.
  Rev.}\ }\textbf {\bibinfo {volume} {103}},\ \bibinfo {pages} {469} (\bibinfo
  {year} {1956})}\BibitemShut {NoStop}%
\bibitem [{\citenamefont {Walecka}(1974)}]{Walecka74}%
  \BibitemOpen
  \bibfield  {author} {\bibinfo {author} {\bibfnamefont {J.~D.}\ \bibnamefont
  {Walecka}},\ }\href@noop {} {\bibfield  {journal} {\bibinfo  {journal} {Ann.
  Phys. (N.Y.)}\ }\textbf {\bibinfo {volume} {83}},\ \bibinfo {pages} {491}
  (\bibinfo {year} {1974})}\BibitemShut {NoStop}%
\bibitem [{\citenamefont {Boguta}\ and\ \citenamefont
  {Bodmer}(1977)}]{Boguta:1977xi}%
  \BibitemOpen
  \bibfield  {author} {\bibinfo {author} {\bibfnamefont {J.}~\bibnamefont
  {Boguta}}\ and\ \bibinfo {author} {\bibfnamefont {A.~R.}\ \bibnamefont
  {Bodmer}},\ }\href {\doibase 10.1016/0375-9474(77)90626-1} {\bibfield
  {journal} {\bibinfo  {journal} {Nucl. Phys.}\ }\textbf {\bibinfo {volume}
  {A292}},\ \bibinfo {pages} {413} (\bibinfo {year} {1977})}\BibitemShut
  {NoStop}%
\bibitem [{\citenamefont {Serot}\ and\ \citenamefont
  {Walecka}(1986)}]{Serot:1984ey}%
  \BibitemOpen
  \bibfield  {author} {\bibinfo {author} {\bibfnamefont {B.~D.}\ \bibnamefont
  {Serot}}\ and\ \bibinfo {author} {\bibfnamefont {J.~D.}\ \bibnamefont
  {Walecka}},\ }\href@noop {} {\bibfield  {journal} {\bibinfo  {journal} {Adv.
  Nucl. Phys.}\ }\textbf {\bibinfo {volume} {16}},\ \bibinfo {pages} {1}
  (\bibinfo {year} {1986})}\BibitemShut {NoStop}%
\bibitem [{\citenamefont {Mueller}\ and\ \citenamefont
  {Serot}(1996)}]{Mueller:1996pm}%
  \BibitemOpen
  \bibfield  {author} {\bibinfo {author} {\bibfnamefont {H.}~\bibnamefont
  {Mueller}}\ and\ \bibinfo {author} {\bibfnamefont {B.~D.}\ \bibnamefont
  {Serot}},\ }\href {\doibase 10.1016/0375-9474(96)00187-X} {\bibfield
  {journal} {\bibinfo  {journal} {Nucl. Phys.}\ }\textbf {\bibinfo {volume}
  {A606}},\ \bibinfo {pages} {508} (\bibinfo {year} {1996})},\ \Eprint
  {http://arxiv.org/abs/nucl-th/9603037} {arXiv:nucl-th/9603037 [nucl-th]}
  \BibitemShut {NoStop}%
\bibitem [{\citenamefont {Typel}\ \emph {et~al.}(2010)\citenamefont {Typel},
  \citenamefont {R{\"o}pke}, \citenamefont {Kl{\"a}hn}, \citenamefont
  {Blaschke},\ and\ \citenamefont {Wolter}}]{Typel:2009sy}%
  \BibitemOpen
  \bibfield  {author} {\bibinfo {author} {\bibfnamefont {S.}~\bibnamefont
  {Typel}}, \bibinfo {author} {\bibfnamefont {G.}~\bibnamefont {R{\"o}pke}},
  \bibinfo {author} {\bibfnamefont {T.}~\bibnamefont {Kl{\"a}hn}}, \bibinfo
  {author} {\bibfnamefont {D.}~\bibnamefont {Blaschke}}, \ and\ \bibinfo
  {author} {\bibfnamefont {H.~H.}\ \bibnamefont {Wolter}},\ }\href@noop {}
  {\bibfield  {journal} {\bibinfo  {journal} {Phys. Rev.}\ }\textbf {\bibinfo
  {volume} {C81}},\ \bibinfo {pages} {015803} (\bibinfo {year} {2010})},\
  \Eprint {http://arxiv.org/abs/0908.2344} {arXiv:0908.2344 [nucl-th]}
  \BibitemShut {NoStop}%
\bibitem [{\citenamefont {Hornick}\ \emph {et~al.}(2018)\citenamefont
  {Hornick}, \citenamefont {Tolos}, \citenamefont {Zacchi}, \citenamefont
  {Christian},\ and\ \citenamefont {Schaffner-Bielich}}]{Hornick:2018kfi}%
  \BibitemOpen
  \bibfield  {author} {\bibinfo {author} {\bibfnamefont {N.}~\bibnamefont
  {Hornick}}, \bibinfo {author} {\bibfnamefont {L.}~\bibnamefont {Tolos}},
  \bibinfo {author} {\bibfnamefont {A.}~\bibnamefont {Zacchi}}, \bibinfo
  {author} {\bibfnamefont {J.-E.}\ \bibnamefont {Christian}}, \ and\ \bibinfo
  {author} {\bibfnamefont {J.}~\bibnamefont {Schaffner-Bielich}},\ }\href
  {\doibase 10.1103/PhysRevC.98.065804} {\bibfield  {journal} {\bibinfo
  {journal} {Phys. Rev.}\ }\textbf {\bibinfo {volume} {C98}},\ \bibinfo {pages}
  {065804} (\bibinfo {year} {2018})},\ \Eprint
  {http://arxiv.org/abs/1808.06808} {arXiv:1808.06808 [astro-ph.HE]}
  \BibitemShut {NoStop}%
\bibitem [{\citenamefont {Abbott}\ \emph {et~al.}(2017)\citenamefont {Abbott}
  \emph {et~al.}}]{TheLIGOScientific:2017qsa}%
  \BibitemOpen
  \bibfield  {author} {\bibinfo {author} {\bibfnamefont {B.~P.}\ \bibnamefont
  {Abbott}} \emph {et~al.} (\bibinfo {collaboration} {Virgo, LIGO
  Scientific}),\ }\href {\doibase 10.1103/PhysRevLett.119.161101} {\bibfield
  {journal} {\bibinfo  {journal} {Phys. Rev. Lett.}\ }\textbf {\bibinfo
  {volume} {119}},\ \bibinfo {pages} {161101} (\bibinfo {year} {2017})},\
  \Eprint {http://arxiv.org/abs/1710.05832} {arXiv:1710.05832 [gr-qc]}
  \BibitemShut {NoStop}%
\bibitem [{\citenamefont {Annala}\ \emph {et~al.}(2018)\citenamefont {Annala},
  \citenamefont {Gorda}, \citenamefont {Kurkela},\ and\ \citenamefont
  {Vuorinen}}]{Annala:2017llu}%
  \BibitemOpen
  \bibfield  {author} {\bibinfo {author} {\bibfnamefont {E.}~\bibnamefont
  {Annala}}, \bibinfo {author} {\bibfnamefont {T.}~\bibnamefont {Gorda}},
  \bibinfo {author} {\bibfnamefont {A.}~\bibnamefont {Kurkela}}, \ and\
  \bibinfo {author} {\bibfnamefont {A.}~\bibnamefont {Vuorinen}},\ }\href
  {\doibase 10.1103/PhysRevLett.120.172703} {\bibfield  {journal} {\bibinfo
  {journal} {Phys. Rev. Lett.}\ }\textbf {\bibinfo {volume} {120}},\ \bibinfo
  {pages} {172703} (\bibinfo {year} {2018})},\ \Eprint
  {http://arxiv.org/abs/1711.02644} {arXiv:1711.02644 [astro-ph.HE]}
  \BibitemShut {NoStop}%
\bibitem [{\citenamefont {Bauswein}\ \emph {et~al.}(2017)\citenamefont
  {Bauswein}, \citenamefont {Just}, \citenamefont {Janka},\ and\ \citenamefont
  {Stergioulas}}]{Bauswein:2017vtn}%
  \BibitemOpen
  \bibfield  {author} {\bibinfo {author} {\bibfnamefont {A.}~\bibnamefont
  {Bauswein}}, \bibinfo {author} {\bibfnamefont {O.}~\bibnamefont {Just}},
  \bibinfo {author} {\bibfnamefont {H.-T.}\ \bibnamefont {Janka}}, \ and\
  \bibinfo {author} {\bibfnamefont {N.}~\bibnamefont {Stergioulas}},\ }\href
  {\doibase 10.3847/2041-8213/aa9994} {\bibfield  {journal} {\bibinfo
  {journal} {Astrophys. J.}\ }\textbf {\bibinfo {volume} {850}},\ \bibinfo
  {pages} {L34} (\bibinfo {year} {2017})},\ \Eprint
  {http://arxiv.org/abs/1710.06843} {arXiv:1710.06843 [astro-ph.HE]}
  \BibitemShut {NoStop}%
\bibitem [{\citenamefont {Paschalidis}\ \emph {et~al.}(2018)\citenamefont
  {Paschalidis}, \citenamefont {Yagi}, \citenamefont {Alvarez-Castillo},
  \citenamefont {Blaschke},\ and\ \citenamefont
  {Sedrakian}}]{Paschalidis:2017qmb}%
  \BibitemOpen
  \bibfield  {author} {\bibinfo {author} {\bibfnamefont {V.}~\bibnamefont
  {Paschalidis}}, \bibinfo {author} {\bibfnamefont {K.}~\bibnamefont {Yagi}},
  \bibinfo {author} {\bibfnamefont {D.}~\bibnamefont {Alvarez-Castillo}},
  \bibinfo {author} {\bibfnamefont {D.~B.}\ \bibnamefont {Blaschke}}, \ and\
  \bibinfo {author} {\bibfnamefont {A.}~\bibnamefont {Sedrakian}},\ }\href
  {\doibase 10.1103/PhysRevD.97.084038} {\bibfield  {journal} {\bibinfo
  {journal} {Phys. Rev.}\ }\textbf {\bibinfo {volume} {D97}},\ \bibinfo {pages}
  {084038} (\bibinfo {year} {2018})},\ \Eprint
  {http://arxiv.org/abs/1712.00451} {arXiv:1712.00451 [astro-ph.HE]}
  \BibitemShut {NoStop}%
\bibitem [{\citenamefont {Alvarez-Castillo}\ \emph {et~al.}(2019)\citenamefont
  {Alvarez-Castillo}, \citenamefont {Blaschke}, \citenamefont {Grunfeld},\ and\
  \citenamefont {Pagura}}]{Alvarez-Castillo:2018pve}%
  \BibitemOpen
  \bibfield  {author} {\bibinfo {author} {\bibfnamefont {D.~E.}\ \bibnamefont
  {Alvarez-Castillo}}, \bibinfo {author} {\bibfnamefont {D.~B.}\ \bibnamefont
  {Blaschke}}, \bibinfo {author} {\bibfnamefont {A.~G.}\ \bibnamefont
  {Grunfeld}}, \ and\ \bibinfo {author} {\bibfnamefont {V.~P.}\ \bibnamefont
  {Pagura}},\ }\href {\doibase 10.1103/PhysRevD.99.063010} {\bibfield
  {journal} {\bibinfo  {journal} {Phys. Rev.}\ }\textbf {\bibinfo {volume}
  {D99}},\ \bibinfo {pages} {063010} (\bibinfo {year} {2019})},\ \Eprint
  {http://arxiv.org/abs/1805.04105} {arXiv:1805.04105 [hep-ph]} \BibitemShut
  {NoStop}%
\bibitem [{\citenamefont {Christian}\ \emph {et~al.}(2019)\citenamefont
  {Christian}, \citenamefont {Zacchi},\ and\ \citenamefont
  {Schaffner-Bielich}}]{Christian:2018jyd}%
  \BibitemOpen
  \bibfield  {author} {\bibinfo {author} {\bibfnamefont {J.-E.}\ \bibnamefont
  {Christian}}, \bibinfo {author} {\bibfnamefont {A.}~\bibnamefont {Zacchi}}, \
  and\ \bibinfo {author} {\bibfnamefont {J.}~\bibnamefont
  {Schaffner-Bielich}},\ }\href {\doibase 10.1103/PhysRevD.99.023009}
  {\bibfield  {journal} {\bibinfo  {journal} {Phys. Rev.}\ }\textbf {\bibinfo
  {volume} {D99}},\ \bibinfo {pages} {023009} (\bibinfo {year} {2019})},\
  \Eprint {http://arxiv.org/abs/1809.03333} {arXiv:1809.03333 [astro-ph.HE]}
  \BibitemShut {NoStop}%
\bibitem [{\citenamefont {Montana}\ \emph {et~al.}(2019)\citenamefont
  {Montana}, \citenamefont {Tolos}, \citenamefont {Hanauske},\ and\
  \citenamefont {Rezzolla}}]{Montana:2018bkb}%
  \BibitemOpen
  \bibfield  {author} {\bibinfo {author} {\bibfnamefont {G.}~\bibnamefont
  {Montana}}, \bibinfo {author} {\bibfnamefont {L.}~\bibnamefont {Tolos}},
  \bibinfo {author} {\bibfnamefont {M.}~\bibnamefont {Hanauske}}, \ and\
  \bibinfo {author} {\bibfnamefont {L.}~\bibnamefont {Rezzolla}},\ }\href@noop
  {} {\bibfield  {journal} {\bibinfo  {journal} {Phys. Rev.}\ }\textbf
  {\bibinfo {volume} {D 99}} (\bibinfo {year} {2019})},\ \Eprint
  {http://arxiv.org/abs/1811.10929} {arXiv:1811.10929 [astro-ph.HE]}
  \BibitemShut {NoStop}%
\bibitem [{\citenamefont {Sieniawska}\ \emph {et~al.}(2019)\citenamefont
  {Sieniawska}, \citenamefont {Turczanski}, \citenamefont {Bejger},\ and\
  \citenamefont {Zdunik}}]{Sieniawska:2018zzj}%
  \BibitemOpen
  \bibfield  {author} {\bibinfo {author} {\bibfnamefont {M.}~\bibnamefont
  {Sieniawska}}, \bibinfo {author} {\bibfnamefont {W.}~\bibnamefont
  {Turczanski}}, \bibinfo {author} {\bibfnamefont {M.}~\bibnamefont {Bejger}},
  \ and\ \bibinfo {author} {\bibfnamefont {J.~L.}\ \bibnamefont {Zdunik}},\
  }\href {\doibase 10.1051/0004-6361/201833969} {\bibfield  {journal} {\bibinfo
   {journal} {Astron. Astrophys.}\ }\textbf {\bibinfo {volume} {622}},\
  \bibinfo {pages} {A174} (\bibinfo {year} {2019})},\ \Eprint
  {http://arxiv.org/abs/1807.11581} {arXiv:1807.11581 [astro-ph.HE]}
  \BibitemShut {NoStop}%
\bibitem [{\citenamefont {Demorest}\ \emph {et~al.}(2010)\citenamefont
  {Demorest}, \citenamefont {Pennucci}, \citenamefont {Ransom}, \citenamefont
  {Roberts},\ and\ \citenamefont {Hessels}}]{Demorest:2010bx}%
  \BibitemOpen
  \bibfield  {author} {\bibinfo {author} {\bibfnamefont {P.}~\bibnamefont
  {Demorest}}, \bibinfo {author} {\bibfnamefont {T.}~\bibnamefont {Pennucci}},
  \bibinfo {author} {\bibfnamefont {S.}~\bibnamefont {Ransom}}, \bibinfo
  {author} {\bibfnamefont {M.}~\bibnamefont {Roberts}}, \ and\ \bibinfo
  {author} {\bibfnamefont {J.}~\bibnamefont {Hessels}},\ }\href {\doibase
  10.1038/nature09466} {\bibfield  {journal} {\bibinfo  {journal} {Nature}\
  }\textbf {\bibinfo {volume} {467}},\ \bibinfo {pages} {1081} (\bibinfo {year}
  {2010})},\ \Eprint {http://arxiv.org/abs/1010.5788} {arXiv:1010.5788
  [astro-ph.HE]} \BibitemShut {NoStop}%
\bibitem [{\citenamefont {Antoniadis}\ \emph {et~al.}(2013)\citenamefont
  {Antoniadis}, \citenamefont {Freire}, \citenamefont {Wex}, \citenamefont
  {Tauris}, \citenamefont {Lynch}, \citenamefont {van Kerkwijk}, \citenamefont
  {Kramer}, \citenamefont {Bassa}, \citenamefont {Dhillon}, \citenamefont
  {Driebe}, \citenamefont {Hessels}, \citenamefont {Kaspi}, \citenamefont
  {Kondratiev}, \citenamefont {Langer}, \citenamefont {Marsh}, \citenamefont
  {McLaughlin}, \citenamefont {Pennucci}, \citenamefont {Ransom}, \citenamefont
  {Stairs}, \citenamefont {van Leeuwen}, \citenamefont {Verbiest},\ and\
  \citenamefont {Whelan}}]{Antoniadis:2013pzd}%
  \BibitemOpen
  \bibfield  {author} {\bibinfo {author} {\bibfnamefont {J.}~\bibnamefont
  {Antoniadis}}, \bibinfo {author} {\bibfnamefont {P.~C.}\ \bibnamefont
  {Freire}}, \bibinfo {author} {\bibfnamefont {N.}~\bibnamefont {Wex}},
  \bibinfo {author} {\bibfnamefont {T.~M.}\ \bibnamefont {Tauris}}, \bibinfo
  {author} {\bibfnamefont {R.~S.}\ \bibnamefont {Lynch}}, \bibinfo {author}
  {\bibfnamefont {M.~H.}\ \bibnamefont {van Kerkwijk}}, \bibinfo {author}
  {\bibfnamefont {M.}~\bibnamefont {Kramer}}, \bibinfo {author} {\bibfnamefont
  {C.}~\bibnamefont {Bassa}}, \bibinfo {author} {\bibfnamefont {V.~S.}\
  \bibnamefont {Dhillon}}, \bibinfo {author} {\bibfnamefont {T.}~\bibnamefont
  {Driebe}}, \bibinfo {author} {\bibfnamefont {J.~W.~T.}\ \bibnamefont
  {Hessels}}, \bibinfo {author} {\bibfnamefont {V.~M.}\ \bibnamefont {Kaspi}},
  \bibinfo {author} {\bibfnamefont {V.~I.}\ \bibnamefont {Kondratiev}},
  \bibinfo {author} {\bibfnamefont {N.}~\bibnamefont {Langer}}, \bibinfo
  {author} {\bibfnamefont {T.~R.}\ \bibnamefont {Marsh}}, \bibinfo {author}
  {\bibfnamefont {M.~A.}\ \bibnamefont {McLaughlin}}, \bibinfo {author}
  {\bibfnamefont {T.~T.}\ \bibnamefont {Pennucci}}, \bibinfo {author}
  {\bibfnamefont {S.~M.}\ \bibnamefont {Ransom}}, \bibinfo {author}
  {\bibfnamefont {I.~H.}\ \bibnamefont {Stairs}}, \bibinfo {author}
  {\bibfnamefont {J.}~\bibnamefont {van Leeuwen}}, \bibinfo {author}
  {\bibfnamefont {J.~P.~W.}\ \bibnamefont {Verbiest}}, \ and\ \bibinfo {author}
  {\bibfnamefont {D.~G.}\ \bibnamefont {Whelan}},\ }\href {\doibase
  10.1126/science.1233232} {\bibfield  {journal} {\bibinfo  {journal}
  {Science}\ }\textbf {\bibinfo {volume} {340}},\ \bibinfo {pages} {6131}
  (\bibinfo {year} {2013})},\ \Eprint {http://arxiv.org/abs/1304.6875}
  {arXiv:1304.6875 [astro-ph.HE]} \BibitemShut {NoStop}%
\bibitem [{\citenamefont {Fonseca}\ \emph {et~al.}(2016)\citenamefont {Fonseca}
  \emph {et~al.}}]{Fonseca:2016tux}%
  \BibitemOpen
  \bibfield  {author} {\bibinfo {author} {\bibfnamefont {E.}~\bibnamefont
  {Fonseca}} \emph {et~al.},\ }\href {\doibase 10.3847/0004-637X/832/2/167}
  {\bibfield  {journal} {\bibinfo  {journal} {Astrophys. J.}\ }\textbf
  {\bibinfo {volume} {832}},\ \bibinfo {pages} {167} (\bibinfo {year}
  {2016})},\ \Eprint {http://arxiv.org/abs/1603.00545} {arXiv:1603.00545
  [astro-ph.HE]} \BibitemShut {NoStop}%
\bibitem [{\citenamefont {Cromartie}\ \emph {et~al.}(2019)\citenamefont
  {Cromartie} \emph {et~al.}}]{Cromartie:2019kug}%
  \BibitemOpen
  \bibfield  {author} {\bibinfo {author} {\bibfnamefont {H.~T.}\ \bibnamefont
  {Cromartie}} \emph {et~al.},\ }\href {\doibase 10.1038/s41550-019-0880-2} {\
  (\bibinfo {year} {2019}),\ 10.1038/s41550-019-0880-2},\ \Eprint
  {http://arxiv.org/abs/1904.06759} {arXiv:1904.06759 [astro-ph.HE]}
  \BibitemShut {NoStop}%
\bibitem [{\citenamefont {{Boguta}}\ and\ \citenamefont
  {{St\"ocker}}(1983)}]{1983PhLB..120..289B}%
  \BibitemOpen
  \bibfield  {author} {\bibinfo {author} {\bibfnamefont {J.}~\bibnamefont
  {{Boguta}}}\ and\ \bibinfo {author} {\bibfnamefont {H.}~\bibnamefont
  {{St\"ocker}}},\ }\href {\doibase 10.1016/0370-2693(83)90446-X} {\bibfield
  {journal} {\bibinfo  {journal} {Physics Letters B}\ }\textbf {\bibinfo
  {volume} {120}},\ \bibinfo {pages} {289} (\bibinfo {year}
  {1983})}\BibitemShut {NoStop}%
\bibitem [{\citenamefont {Yasin}\ \emph {et~al.}(2018)\citenamefont {Yasin},
  \citenamefont {Sch\"afer}, \citenamefont {Arcones},\ and\ \citenamefont
  {Schwenk}}]{Yasin:2018ckc}%
  \BibitemOpen
  \bibfield  {author} {\bibinfo {author} {\bibfnamefont {H.}~\bibnamefont
  {Yasin}}, \bibinfo {author} {\bibfnamefont {S.}~\bibnamefont {Sch\"afer}},
  \bibinfo {author} {\bibfnamefont {A.}~\bibnamefont {Arcones}}, \ and\
  \bibinfo {author} {\bibfnamefont {A.}~\bibnamefont {Schwenk}},\ }\href@noop
  {} {\  (\bibinfo {year} {2018})},\ \Eprint {http://arxiv.org/abs/1812.02002}
  {arXiv:1812.02002 [nucl-th]} \BibitemShut {NoStop}%
\bibitem [{\citenamefont {Alford}\ \emph {et~al.}(2014)\citenamefont {Alford},
  \citenamefont {Han},\ and\ \citenamefont {Prakash}}]{Alford:2014dva}%
  \BibitemOpen
  \bibfield  {author} {\bibinfo {author} {\bibfnamefont {M.~G.}\ \bibnamefont
  {Alford}}, \bibinfo {author} {\bibfnamefont {S.}~\bibnamefont {Han}}, \ and\
  \bibinfo {author} {\bibfnamefont {M.}~\bibnamefont {Prakash}},\ }\href
  {\doibase 10.7566/JPSCP.1.013041} {\bibfield  {journal} {\bibinfo  {journal}
  {JPS Conf.Proc.}\ }\textbf {\bibinfo {volume} {1}},\ \bibinfo {pages}
  {013041} (\bibinfo {year} {2014})}\BibitemShut {NoStop}%
\bibitem [{\citenamefont {Riley}\ \emph {et~al.}(2019)\citenamefont {Riley}
  \emph {et~al.}}]{Riley:2019yda}%
  \BibitemOpen
  \bibfield  {author} {\bibinfo {author} {\bibfnamefont {T.~E.}\ \bibnamefont
  {Riley}} \emph {et~al.},\ }\href {\doibase 10.3847/2041-8213/ab481c}
  {\bibfield  {journal} {\bibinfo  {journal} {Astrophys. J. Lett.}\ }\textbf
  {\bibinfo {volume} {887}},\ \bibinfo {pages} {L21} (\bibinfo {year}
  {2019})},\ \Eprint {http://arxiv.org/abs/1912.05702} {arXiv:1912.05702
  [astro-ph.HE]} \BibitemShut {NoStop}%
\bibitem [{\citenamefont {Miller}\ \emph {et~al.}(2019)\citenamefont {Miller}
  \emph {et~al.}}]{Miller:2019cac}%
  \BibitemOpen
  \bibfield  {author} {\bibinfo {author} {\bibfnamefont {M.~C.}\ \bibnamefont
  {Miller}} \emph {et~al.},\ }\href@noop {} {\bibfield  {journal} {\bibinfo
  {journal} {Astrophys. J. Lett.}\ }\textbf {\bibinfo {volume} {887}},\
  \bibinfo {pages} {L24} (\bibinfo {year} {2019})},\ \Eprint
  {http://arxiv.org/abs/1912.05705} {arXiv:1912.05705 [astro-ph.HE]}
  \BibitemShut {NoStop}%
\bibitem [{\citenamefont {Raaijmakers}\ \emph {et~al.}(2019)\citenamefont
  {Raaijmakers} \emph {et~al.}}]{Raaijmakers:2019qny}%
  \BibitemOpen
  \bibfield  {author} {\bibinfo {author} {\bibfnamefont {G.}~\bibnamefont
  {Raaijmakers}} \emph {et~al.},\ }\href {\doibase 10.3847/2041-8213/ab451a}
  {\bibfield  {journal} {\bibinfo  {journal} {Astrophys. J. Lett.}\ }\textbf
  {\bibinfo {volume} {887}},\ \bibinfo {pages} {L22} (\bibinfo {year}
  {2019})},\ \Eprint {http://arxiv.org/abs/1912.05703} {arXiv:1912.05703
  [astro-ph.HE]} \BibitemShut {NoStop}%
\bibitem [{\citenamefont {Todd-Rutel}\ and\ \citenamefont
  {Piekarewicz}(2005)}]{ToddRutel:2005fa}%
  \BibitemOpen
  \bibfield  {author} {\bibinfo {author} {\bibfnamefont {B.~G.}\ \bibnamefont
  {Todd-Rutel}}\ and\ \bibinfo {author} {\bibfnamefont {J.}~\bibnamefont
  {Piekarewicz}},\ }\href {\doibase 10.1103/PhysRevLett.95.122501} {\bibfield
  {journal} {\bibinfo  {journal} {Phys. Rev. Lett.}\ }\textbf {\bibinfo
  {volume} {95}},\ \bibinfo {pages} {122501} (\bibinfo {year} {2005})},\
  \Eprint {http://arxiv.org/abs/nucl-th/0504034} {arXiv:nucl-th/0504034
  [nucl-th]} \BibitemShut {NoStop}%
\bibitem [{\citenamefont {Chen}\ and\ \citenamefont
  {Piekarewicz}(2014)}]{Chen:2014sca}%
  \BibitemOpen
  \bibfield  {author} {\bibinfo {author} {\bibfnamefont {W.-C.}\ \bibnamefont
  {Chen}}\ and\ \bibinfo {author} {\bibfnamefont {J.}~\bibnamefont
  {Piekarewicz}},\ }\href {\doibase 10.1103/PhysRevC.90.044305} {\bibfield
  {journal} {\bibinfo  {journal} {Phys. Rev.}\ }\textbf {\bibinfo {volume}
  {C90}},\ \bibinfo {pages} {044305} (\bibinfo {year} {2014})},\ \Eprint
  {http://arxiv.org/abs/1408.4159} {arXiv:1408.4159 [nucl-th]} \BibitemShut
  {NoStop}%
\bibitem [{\citenamefont {Horowitz}\ and\ \citenamefont
  {Piekarewicz}(2001)}]{Horowitz:2001ya}%
  \BibitemOpen
  \bibfield  {author} {\bibinfo {author} {\bibfnamefont {C.~J.}\ \bibnamefont
  {Horowitz}}\ and\ \bibinfo {author} {\bibfnamefont {J.}~\bibnamefont
  {Piekarewicz}},\ }\href@noop {} {\bibfield  {journal} {\bibinfo  {journal}
  {Phys. Rev.}\ }\textbf {\bibinfo {volume} {C64}},\ \bibinfo {pages} {062802}
  (\bibinfo {year} {2001})},\ \Eprint {http://arxiv.org/abs/nucl-th/0108036}
  {nucl-th/0108036} \BibitemShut {NoStop}%
\bibitem [{\citenamefont {Drischler}\ \emph {et~al.}(2016)\citenamefont
  {Drischler}, \citenamefont {Carbone}, \citenamefont {Hebeler},\ and\
  \citenamefont {Schwenk}}]{Drischler:2016djf}%
  \BibitemOpen
  \bibfield  {author} {\bibinfo {author} {\bibfnamefont {C.}~\bibnamefont
  {Drischler}}, \bibinfo {author} {\bibfnamefont {A.}~\bibnamefont {Carbone}},
  \bibinfo {author} {\bibfnamefont {K.}~\bibnamefont {Hebeler}}, \ and\
  \bibinfo {author} {\bibfnamefont {A.}~\bibnamefont {Schwenk}},\ }\href
  {\doibase 10.1103/PhysRevC.94.054307} {\bibfield  {journal} {\bibinfo
  {journal} {Phys. Rev.}\ }\textbf {\bibinfo {volume} {C94}},\ \bibinfo {pages}
  {054307} (\bibinfo {year} {2016})},\ \Eprint
  {http://arxiv.org/abs/1608.05615} {arXiv:1608.05615 [nucl-th]} \BibitemShut
  {NoStop}%
\bibitem [{\citenamefont {Alford}\ and\ \citenamefont
  {Han}(2016)}]{Alford:2015gna}%
  \BibitemOpen
  \bibfield  {author} {\bibinfo {author} {\bibfnamefont {M.~G.}\ \bibnamefont
  {Alford}}\ and\ \bibinfo {author} {\bibfnamefont {S.}~\bibnamefont {Han}},\
  }\href {\doibase 10.1140/epja/i2016-16062-9} {\bibfield  {journal} {\bibinfo
  {journal} {Eur. Phys. J.}\ }\textbf {\bibinfo {volume} {A52}},\ \bibinfo
  {pages} {62} (\bibinfo {year} {2016})},\ \Eprint
  {http://arxiv.org/abs/1508.01261} {arXiv:1508.01261 [nucl-th]} \BibitemShut
  {NoStop}%
\bibitem [{\citenamefont {Abbott}\ \emph {et~al.}(2019)\citenamefont {Abbott}
  \emph {et~al.}}]{Abbott:2018wiz}%
  \BibitemOpen
  \bibfield  {author} {\bibinfo {author} {\bibfnamefont {B.~P.}\ \bibnamefont
  {Abbott}} \emph {et~al.} (\bibinfo {collaboration} {LIGO Scientific,
  Virgo}),\ }\href {\doibase 10.1103/PhysRevX.9.011001} {\bibfield  {journal}
  {\bibinfo  {journal} {Phys. Rev.}\ }\textbf {\bibinfo {volume} {X9}},\
  \bibinfo {pages} {011001} (\bibinfo {year} {2019})},\ \Eprint
  {http://arxiv.org/abs/1805.11579} {arXiv:1805.11579 [gr-qc]} \BibitemShut
  {NoStop}%
\bibitem [{\citenamefont {Hinderer}(2008)}]{Hinderer:2007mb}%
  \BibitemOpen
  \bibfield  {author} {\bibinfo {author} {\bibfnamefont {T.}~\bibnamefont
  {Hinderer}},\ }\href {\doibase 10.1086/533487} {\bibfield  {journal}
  {\bibinfo  {journal} {Astrophys. J.}\ }\textbf {\bibinfo {volume} {677}},\
  \bibinfo {pages} {1216} (\bibinfo {year} {2008})},\ \Eprint
  {http://arxiv.org/abs/0711.2420} {arXiv:0711.2420 [astro-ph]} \BibitemShut
  {NoStop}%
\bibitem [{\citenamefont {Hinderer}\ \emph {et~al.}(2010)\citenamefont
  {Hinderer}, \citenamefont {Lackey}, \citenamefont {Lang},\ and\ \citenamefont
  {Read}}]{Hinderer:2009ca}%
  \BibitemOpen
  \bibfield  {author} {\bibinfo {author} {\bibfnamefont {T.}~\bibnamefont
  {Hinderer}}, \bibinfo {author} {\bibfnamefont {B.~D.}\ \bibnamefont
  {Lackey}}, \bibinfo {author} {\bibfnamefont {R.~N.}\ \bibnamefont {Lang}}, \
  and\ \bibinfo {author} {\bibfnamefont {J.~S.}\ \bibnamefont {Read}},\ }\href
  {\doibase 10.1103/PhysRevD.81.123016} {\bibfield  {journal} {\bibinfo
  {journal} {Phys. Rev.}\ }\textbf {\bibinfo {volume} {D81}},\ \bibinfo {pages}
  {123016} (\bibinfo {year} {2010})},\ \Eprint {http://arxiv.org/abs/0911.3535}
  {arXiv:0911.3535 [astro-ph.HE]} \BibitemShut {NoStop}%
\bibitem [{\citenamefont {Postnikov}\ \emph {et~al.}(2010)\citenamefont
  {Postnikov}, \citenamefont {Prakash},\ and\ \citenamefont
  {Lattimer}}]{Postnikov:2010yn}%
  \BibitemOpen
  \bibfield  {author} {\bibinfo {author} {\bibfnamefont {S.}~\bibnamefont
  {Postnikov}}, \bibinfo {author} {\bibfnamefont {M.}~\bibnamefont {Prakash}},
  \ and\ \bibinfo {author} {\bibfnamefont {J.~M.}\ \bibnamefont {Lattimer}},\
  }\href {\doibase 10.1103/PhysRevD.82.024016} {\bibfield  {journal} {\bibinfo
  {journal} {Phys. Rev.}\ }\textbf {\bibinfo {volume} {D82}},\ \bibinfo {pages}
  {024016} (\bibinfo {year} {2010})},\ \Eprint {http://arxiv.org/abs/1004.5098}
  {arXiv:1004.5098 [astro-ph.SR]} \BibitemShut {NoStop}%
\end{thebibliography}%
\end{document}